\documentclass{pasa}

\title[Comparison of classification techniques for masers]{Comparison of three Statistical Classification Techniques for Maser Identification}
\author[Manning et al.]{Ellen\ M. Manning,$^{1}$ Barbara\ R. Holland,$^{1}$ Simon\ P. Ellingsen,$^{1}$ Shari\ L. Breen,$^{2}$ Xi\ Chen,$^{3,4}$ Melissa\ Humphries$^{1}$\\
\affil{$^1$School of Physical Sciences, University of Tasmania, Private Bag 37, Hobart, Tasmania 7001, Australia}
\affil{$^2$CSIRO Asronomy and Space Science, PO Box 76, Epping, NSW 1710, Australia}
\affil{$^3$Key Laboratory for Research in Galaxies and Cosmology, Shanghai Astronomical Observatory, Chinese Academy of Sciences, Shanghai 200030, China}
\affil{$^4$Key Laboratory of Radio Astronomy, Chinese Academy of Sciences, China}}
\jid{PASA}
\doi{10.1017/pas.\the\year.xxx}
\jyear{\the\year}

\usepackage[authoryear]{natbib}
\bibpunct{(}{)}{;}{a}{}{,}
\usepackage{aas_macros}
\usepackage{hyperref} 
\hypersetup{colorlinks,citecolor=blue,linkcolor=blue,urlcolor=blue}

\usepackage{floatrow}
\DeclareFloatFont{tiny}{\tiny}
\usepackage{epsfig}
\usepackage{epstopdf}
\usepackage{graphicx}
\usepackage{color}
\usepackage{fancyvrb}
\usepackage{multirow,multicol}

\newcommand{\ionhy}{H{\sc ii} }

\begin{document}

\VerbatimFootnotes

\begin{abstract}
 We applied three statistical classification techniques - linear discriminant analysis (LDA), logistic regression and random forests - to three astronomical datasets associated with searches for interstellar masers.  We compared the performance of these methods in identifying whether specific mid-infrared or millimetre continuum sources are likely to have associated interstellar masers. We also discuss the ease, or otherwise, with which the results of each classification technique can be interpreted.  Non-parametric methods have the potential to make accurate predictions when there are complex relationships between critical parameters.  We found that for the small datasets the parametric methods logistic regression and LDA performed best, for the largest dataset the non-parametric method of random forests performed with comparable accuracy to parametric techniques, rather than any significant improvement.  This suggests that at least for the specific examples investigated here accuracy of the predictions obtained is not being limited by the use of parametric models. We also found that for LDA, transformation of the data to match a normal distribution in the input parameters led to big improvements in accuracy.  The different classification techniques had significant overlap in their predictions, further astronomical observations will enable the accuracy of these predictions to be tested.

\end{abstract}


\maketitle

\section{Introduction}

In recent years astronomical instrumentation across a range of wavelength bands has improved to the point where high-resolution, sensitive surveys of large areas of the sky are becoming much more common \citep[e.g.][]{Benjamin+03,Johnston+07}.  The higher data rates from new instrumentation and large surveys give the opportunity to collect detailed information on very large numbers of sources and undertake more sophisticated statistical investigations of their properties. This will enable both more reliable identification of sub-groups within the broader population, and identification of rare or unusual objects. However, these new instruments also present the astronomical community with a challenge of how best to extract the maximum utility from large volumes of data.

The desire to accurately and efficiently classify astronomical sources identified in large surveys into different groups is an increasingly common one.  Attempts to develop efficient criteria for targeted searches for interstellar masers, is one specific example of an application of survey source classification.  A number of studies have found that star formation regions with an associated interstellar maser differ significantly in their infrared or millimetre continuum properties from the majority of the population \citep[e.g.][]{Ellingsen05,Ellingsen06,Chen+11}.  In developing criteria for targeting future searches it is desirable to identify a large fraction of the population of interest while including only a small number of sources which do not yield detections.  In the terminology of classification it is important to minimise both the number of false-negatives and false-positives.  A related issue is in understanding the characteristics through which the classification has been achieved.  For example, if you are able to develop efficient criteria for targeting a search for interstellar masers on the basis of infrared or millimetre continuum properties, what is the physical meaning of those characteristics - do they correspond to a particular mass range, or evolutionary phase of the associated high-mass star formation region?

Maser emission occurs naturally in a range of astrophysical environments, including the molecular gas close to newly forming stars, the envelopes of late-type stars, and close to the nuclei of some active galaxies.  Masers have proven to be a reliable signpost of the very early stages of high-mass star formation \citep[e.g.][]{Ellingsen06}; with recent improvements in the availability of sensitive large-area surveys at mid-infrared through millimetre wavelengths, they are increasingly being used as tools to study high-mass star formation \citep[e.g.][]{Titmarsh+13}.  Masers can provide information on the dynamics of the star formation region through observations of their kinematics \citep[e.g.][]{Goddi+11}, on the magnetic field from observations of the polarisation \citep[e.g.][]{Surcis+12}, and potentially the presence and absence of different transitions can provide an evolutionary timeline \citep[e.g.][]{Ellingsen+07,Breen+10a}.  If it is possible to use  classification techniques to reliably identify which regions host different types of maser transition, then an understanding of the physical properties of those regions in combination with the maser-based evolutionary timeline could provide important insights into the formation of high-mass stars.

These types of classification problems are commonly encountered in a wide range of scientific disciplines, and from the broader literature we have been able to identify a number of commonly used classification techniques.  When considering different classification methods, \citet{Breiman+01a} has suggested that there is a trade off between parametric techniques that are easy to interpret but not always as accurate, and non-parametric methods that are more difficult to interpret, but deliver a higher level of accuracy.  Here, we use three different classification techniques to investigate their strengths and weaknesses when applied to the specific problem of efficiently identifying target sources for searches for interstellar masers.  The three methods we have chosen for our investigation are linear discriminant analysis (LDA), logistic regression, and random forests. These three methods were chosen because they have proven effective across a wide range of problem domains, they are relatively easy to implement, and they include two parametric methods (LDA and logistic regression) and one non-parametric method (random forests).

LDA uses similar calculations and techniques to principal component analysis (PCA) which is quite widely used in astronomy \citep[e.g.][]{Lo+09, Einasto+11}. \citet{Kobel+09} used LDA in their classification of different photospheric magnetic elements on the Sun.  They found that the predictions they were able to make on the basis of LDA showed good agreement with the results from previous studies. This can in part be credited to the semi-artificial segregation between the classes of photospheric magnetic elements, as the variables chosen were those with the most significant differences in brightness values. Logistic regression has been less commonly applied in astronomy than PCA, although it has previously been used to successfully  identify which star formation regions are more likely to host different types of interstellar masers \citep[e.g.][]{Breen+07,Ellingsen+10}.  \citet{Yuan+10} and \citet{Song+09} have both shown that logistic regression can be an effective means of predicting solar flares.  Random forests are a relatively new, non-parametric classification technique which has proven to be very effective in other fields, such as ecology.  \citet{Cutler+07} compared the results of classifying ecological data, using the same classification methods as are used here and found that random forests had the highest accuracy.  Within astronomy, random forests have been used by \citet{Bailey+07} to improve the reliability of finding supernovae from images, while \citet{Carliles+10} used them to assign photometric redshifts.  Recently they have also been used as the basis of processes for automated rapid classification and decision making.  \citet{Morgan+12} used random forests as part of a method for making time-efficient recommendations as to which gamma-ray burst events are likely to be high-redshift in order to prioritise whether a specific event deserves additional observing time. They found that by observing the top 20  \% of recommended events it was possible to identify 56  \% of the high-redshift bursts, while using the top 40  \% of recommendations allows identification of 84  \% of high-redshift events.  \citet{Mirabal+12} used random forests to accurately classify whether unidentified objects detected in Gamma-rays by the {\em Fermi} satellite were likely to be Active Galactic Nuclei (AGN) or pulsars (they achieved accuracies of 97.7 and 96.5  \% for AGN and pulsar identification, respectively).

To better understand the strengths and limitations of these different  classification techniques, both in terms of their efficiency and the degree to which the outcomes of the classification process can be related to the properties of the astronomical sources, we compared their performance on three published datasets \citep{Breen+07,Ellingsen+10,Chen+12}. For each of these three sets of data we applied the three classification techniques to make predictions as to which infrared (or millimetre) sources are likely to also be associated with masers.  In Section~\ref{sec:techniques} we describe in more detail each of the  classification techniques used. The properties of each of the datasets are outlined in Section~\ref{sec:results} where we examine the results of applying the different classification techniques in each case. 

\section{Classification Techniques} \label{sec:techniques}

In the context of the current work our data typically consists of astronomical sources for which a range of parameters (e.g. the intensity in a particular wavelength range) have been measured, along with parameters which are related to the quality or uncertainty in the measurement and others which identify the particular astronomical object (e.g. the source number or coordinates).  These parameters are all potential inputs to the different  classification techniques and we refer to these as predictor variables.  In the field of machine learning these are often referred to as features, however, as that term frequently has a different meaning in astronomical literature we do not use that terminology here.  For some (sometimes all),  of the sources in the data set we also have information as to whether or not that source has an associated maser emission from a specific molecular transition.  Hence, we are seeking to accurately classify our astronomical sources into two classes, those with an associated interstellar maser and those without.

\subsection{Linear Discriminant Analysis} \label{sec:LDA}


Linear discriminant analysis finds the  linear combination of predictor variables which maximises the separation of the different classes and minimises the variation within classes \citep{Feigelson+12}.  LDA can be visualised geometrically as projection from a high-dimensional space onto a line. When given a new source to classify, LDA uses this linear combination to convert the high-dimensional data to a real number, and the classification of the sample is determined by comparing this number to a threshold value.  The technique is relatively simple and so is unsuitable if there are complex, nonlinear interactions between the variables.  LDA is a technique of dimensionality reduction similar to principal component analysis (PCA), which is more commonly used in astronomy.  Both LDA and PCA attempt to model the data with linear combinations of the predictor variables; the difference is that PCA does not use classification information in producing the model, whereas LDA does \citep{Feigelson+12}. 

The assumptions of LDA are that the data follows a multivariate normal distribution for each class, classes may have different means but are assumed to have the same variance structure. This makes LDA a parametric method in the sense that it assumes a particular model of the data.  Most astronomical data are not normally distributed, so transformations of the variables are usually required.  For each of the three datasets we studied, LDA was applied to both the original data and to the transformed data as a comparison.  Data Set 2 required an inverse function to normalise the data (each predictor variable was transformed via a $\frac{1}{x}$ function).  In the case of Data Sets 1 and 3 where the samples were naturally clustered, an inverse transformation would have destroyed the bimodality present.  For this reason, a log transformation was selected as it improved the normality of the data while still being easy to interpret.

LDA models were fitted using the {\tt lda} function in $R$ \citep[][part of the {\tt MASS} package]{R}, we left the {\tt prior} input parameter at its default setting  which is to assume that probability of being in a particular class is equal to the relative frequency of the class in the training data.

\subsection{Logistic Regression}  

Logistic regression is a form of generalised linear modelling that is used to predict the probability of an event occurring; in this case, whether or not an astrophysical source has an associated interstellar maser. The probability of occurrence $P$ is calculated from 
\[
P = \frac{1}{1+e^{-z}},
\label{eqn:logistic}
\]
where $z = b_0 + b_1.x_1 + ... + b_n.x_n$, the $b$ values are regression coefficients and the $x_i$ values are the predictor variables. $P$ is then compared to a cut-off threshold of 0.5 (50\% likelihood) to determine whether an object is predicted to have an associated maser, or not.  Like LDA, logistic regression is also a parametric method.

Linear regression assumes that the response variable is normally distributed, in contrast logistic regression assumes that the response variable follows a binomial distribution (which is applicable in our case of two classes).  This means that the method of least squares (used in linear regression), cannot be applied to logistic regression \citep{Hosmer+00}.  Instead, maximum likelihood \citet[formulated by][]{Fisher+22} is used to estimate the parameters of the model. The likelihood function is calculated using the product of contributions to the model from each of the predictor variables \citep{Hosmer+00}.   

Logistic regression was implemented using the function {\tt glm} which is part of the base $R$ package  \citep{R}. To perform a logistic regression the family option in {\tt glm} is set to {\em binomial} and the link function is set to {\em logit}.  It was not feasible to alter any other input parameters in the function to produce our models.

\subsection{Random Forests} 

Classification trees are a non-parametric technique of classification (in contrast to both logistic regression and LDA), which means that they do not assume an underlying model of the data \citep{Cutler+07,Carliles+10}. Classification trees can be more accurate than parametric approaches when complex interactions occur between the predictor variables. This could be the expected case for maser association with infrared or millimetre sources, as well as a broad range of astronomical classification problems. Individual classification trees may not be very accurate, especially when there are more than a few predictor variables, however, a collection of trees grown independently on randomly perturbed versions of the data greatly increases the accuracy of predictions \citep{Breiman+01c,Cutler+07}.  Random forests work by producing large numbers of classification trees and then determining the classification of a particular sample (in our case, an astronomical source) by allowing each of these trees to ``vote'' and then taking the majority rule \citep{Breiman+01b}.  This voting system is also how the probability of a sample being classified into a certain group is calculated; by dividing the number of trees voting for a certain classification by the total number of trees. 

To produce individual classification trees in a random forest a bootstrap sample  is selected for each tree.  For a data set with $N$ entries, $N$ samples are taken. Because sampling is done with replacement, approximately two thirds of the original data occurs at least once in each bootstrap sample \citep{Efron+94}. \citet{Hastie+01} showed that bootstrap sampling causes the variance of the estimated class to converge to a lower limit when more trees are added to the forest, and so rarely overfit \citep{Breiman+01c}.  A classification tree is grown from each bootstrap sample using recursive binary partitioning. The branching points of the trees are called nodes.  In standard trees, the predictor variable at each node is chosen based on the best split, which is determined by the Gini index \citep[a measure of statistical dispersion, see][pg 271]{Hastie+01}.  In a random forest the variable providing the best split is chosen from a random subset of predictor variables \citep{Liaw+02}. The predictor variable and the subset of predictor variables from which it is chosen is independent of any other nodes' variable choices.  This approach decreases the dependence between individual trees.  The splitting process continues until further subdivision no longer decreases the Gini index.  The final classification given by each tree depends on the terminal node the source has been allocated to.

A nice feature of random forests is that they have an inbuilt way of estimating the classification error because of the use of bootstrapping to select slightly different data for each tree. Data not included in the bootstrap sample (approximately one third of observations) for a particular tree are referred to as out-of-bag (oob) values.  The tree grown from each bootstrap sample is used to predict the classification for each of the oob values, giving an estimate of the classification error as well as a means to compare the importance of each variable in the classification process \citep{Breiman}.  The importance of a variable is expressed by the difference between the probability of predicting the class correctly in shuffled oob data (the sample order is rearranged to eliminate systematic errors) compared to the unshuffled oob data \citep{Cutler+07}.  

Random forests also give a natural metric for determining the similarity of two different astronomical sources (or other groups of samples). Proximities between two sources are calculated in the random forest process.  If a pair of sources end up in the same terminal node, their proximity is increased by one.  Similar source pairs end up in the same terminal node more often than dissimilar ones.  The proximities are then normalised (divided by the total number of trees) and the proximity of a point and itself is set to be one.  The proximities are then expressed as a symmetric matrix, where the diagonal entries all have the value one.  The proximity matrix can be used as input for multidimensional scaling, as a way of visualising the classification results (displayed in Sec. \ref{sec:results}).

A potential drawback of random forests is that they cannot be used to directly test hypotheses \citep{Cutler+07}.  They also do not give a clear representation of the actual classification process.  However, although the internal calculations are difficult to interpret, they produce useful properties such as relative variable importance and an estimate of the classification error without extra external calculations \citep{Breiman}.

To create the random forests used in the modelling and classification, we used the $R$ function {\tt randomForest} (in the {\tt randomForest} package).  For an introduction to the usage and features of {\tt randomForest} functions in the $R$ environment, see \citet{Liaw+02}.  There are a number of parameters that can be varied when growing the random forest in order to optimise its classification and predictive accuracy.  These include the number of trees in the forest, the number of variables randomly sampled as candidates at each split, and the maximum number of terminal nodes in the trees.  The minimum size of the terminal nodes can also be varied, where a larger number leads to smaller trees which take less time to grow.  Setting the node size to $k$ means that no node with fewer than $k$ cases will be split \citep{Breiman}.  A terminal node size of 1 is therefore the most accurate, but in cases with large datasets, memory constraints may require this to be higher.  We found that altering these parameters did not consistently increase the sensitivity or specificity significantly, so the default values for the parameters were used: 500 trees grown in the forest, a node size of 1 (default for regression is 5), and the maximum possible number of terminal nodes.  The default number of variables chosen at each split is $\sqrt{p}$ for classification and $p/3$ for regression (rounded to the nearest integer), where $p$ is the total number of predictor variables in the data set.  Other factors that can be varied are whether or not the cases are sampled with replacement (the default, which we used, is with replacement), and the prior probability of each class occurring can also be set with the default being to assume equal class probabilities.

For both Data Set 2 and 3 (where predictions were done), random forests were grown using 3000 trees rather than the default 500.  Since each tree is grown independently, this is equivalent to combining the results of multiple smaller forests.  3000 trees was chosen for both data sets because this produced the most accurate results in the cross validation.  Generally random forests is robust against over-fitting \citep[see][]{Breiman+01c}, however in the case of Data Set 2, due to the very small training set compared to its number of predictor variables, more than 3000 trees decreased the classification accuracy.  In the case of Data Set 3, using more than 3000 trees had no effect.

\subsection{Accuracy of classification techniques} \label{sec:classification}

There are four possible outcomes of the classification of each astronomical source.  The two desired outcomes are  that the classification technique can  correctly identify a source which does have an associated maser (a ``true positive''), or it can correctly identify a source as not having a maser (a ``true negative'').   A perfect classification would have all samples with one or the other of these outcomes.  There are however, two ways in which the classification scheme can give an incorrect outcome and depending on the circumstances these are not necessarily of equal importance.  A ``false positive'' outcome is where a source which does not have an associated maser is classified as being associated with one, while a ``false negative'' occurs when a source which does have an associated maser is classified as not having one associated (see the Confusion Matrix in Table~\ref{tab:Conf}).

For each classification method we calculated both the sensitivity (known as recall in machine learning) and specificity.  In this context, the sensitivity, or true positive rate (TPR), is the  percentage of maser associations correctly predicted by the model, and specificity is the  percentage of maser non-associations correctly predicted, or the true negative rate (TNR).

\[
\mbox{Sensitivity (TPR)} = \frac{\mbox{True Positives}}{\mbox{True Postives + False Negatives}}
\]

\[
\mbox{Specificity (TNR)} = \frac{\mbox{True Negatives}}{\mbox{True Negatives + False Positives}}
\]

\begin{table}
\centering
\begin{tabular}{c|cc}
  & \textbf{Know Negatives} & \textbf{Known Positives}   \\ \hline
\textbf{Classified as} &  \multirow{ 2}{*}{True Negatives} & \multirow{ 2}{*}{False Negatives} \\
\textbf{Negative} & &  \\ 
 & & \\
\textbf{Classified as} & \multirow{ 2}{*}{False Positive} & \multirow{ 2}{*}{True Positive} \\
\textbf{Positive} & &  \\ \hline
 & Specificity & Sensitivity \\
 & (True Negative Rate) & (True Positive Rate) \\
 & & \\
\end{tabular}
\caption{The relationship between the four possible classification results and the calculated values of the sensitivity and the specificity.}
\label{tab:Conf}
\end{table}

\subsubsection{Predictor Variable Importance} \label{sec:var}

Logistic regression performed using $R$ has two techniques for determining the importance of the variables included in the model.  The first is a set of P-values provided when the logistic regression is performed.  The second is the in-built {\tt stepAIC} function, which includes all possible predictor variables in the starting model and iteratively removes variables which do not significantly contribute to the model to yield the most parsimonious model with the greatest predictive power.  To determine which variables to include in the logistic regression models, we used a combination of the {\tt stepAIC} function and manual variable selection.  Variables that did not  increase the accuracy of the model were excluded (see Sec.~\ref{sec:results}).

For LDA variable selection was done manually.  We used the logistic regression's selection as a starting point, and then included additional predictors if they improved the prediction accuracy.

Random forests includes an internal calculation of the Mean Decrease in Accuracy for each of the variables utilised, which is a measure of how poorly the model performs when that variable is not included.   Thus, the higher the value is, the more the predictor variable contributes to the accuracy of the model.  Negative values decrease the accuracy and values close to zero offer little or no effect.  It is worth noting that random forests is potentially robust enough to deal with all available variables and so including them all in the model does not generally decrease the accuracy significantly \citep{Feigelson+12}.

\subsubsection{Cross Validation} 
\label{sec:CV}

The aim of classification is to build models that will generalise well to new data. When constructing models there is a danger in over-fitting to the training data. In order to determine the accuracy of each of the classification methods on the three data sets, we used a 10-fold cross validation technique. Using a fitted model that has been trained on a randomly chosen 90\% of the data, the classification of the remaining tenth is predicted. This procedure of training and prediction is then repeated 1000 times in order to obtain an estimate of the classification error. Repeating the cross validation ensures that a high number of the possible combinations of the data are used, reducing sampling bias associated with randomly folding the data. Repeated 10-fold cross validation of this kind is especially useful when modelling a random forest as the over-fitting associated with regression tree techniques is compensated for by the generous error estimation of the cross validation \citep{Borra+10}. 

In repeated 10-fold cross validation, the results from the multiple runs are averaged. In this case the averaged cross validation produced a mean probability of being associated with a maser for each sample. A source was classified as a maser if the probability was 50\% or above. The percentage of predicted classifications were then compared to the actual classifications (maser source or non-maser source) to determine the accuracy for each model for each of the three data sets.  Adjusting the cut-off threshold for maser classification from 50\% was also investigated to explore the trade-off between sensitivity and specificity of the model. This is useful information to have available when it is important to obtain all the positive classifications, even when it means many false positives are given, and alternatively the model can be adjusted so that there is only a very small chance of a false positive, at the expense of false negative classifications. The receiver operating characteristic (ROC) curves (explained in Sec.~\ref{sec:ROC}), display the results of this analysis.

\subsubsection{Receiver Operating Characteristic Curves} \label{sec:ROC}

A ROC curve plots the true positive rate (sensitivity) against the false positive rate (1 $-$ specificity), effectively showing the trade-off in prediction power for accuracy in a given classification model.  The diagonal line $y = x$ represents randomly classifying the samples, with half predicted as positive and half as negative.  Anywhere in the space above this line means that the model is better than random classification, with the best possible system showing 100\% sensitivity with no false predictions, resulting in a point in the top left hand corner.  ROC curves were plotted here to compare each classification method for each data set in Figures~\ref{fig:ROC-D}, \ref{fig:ROC-S} and~\ref{fig:ROC-B}.

\section{Results} \label{sec:results}

\subsection{Water Masers associated with Star formation regions in the RCW106 Giant Molecular Cloud}
\label{sec:dust}

\citet{Breen+07} undertook a complete search for 22~GHz water masers within the giant molecular cloud RCW~106. This search detected nine 22~GHz water masers and the region searched included 73 1.2-mm dust clumps observed and characterised by \citet{Mookerjea+04}.  Seven of the dust clumps were found to be associated with masers \citep{Breen+07}. \citeauthor{Breen+07} used a form of logistic regression called binomial generalised linear modelling (GLM) to investigate the properties of the astronomical sources (in this case dust clumps) with and without water masers in RCW106. They found that water masers were associated with those sources which are denser, more massive and have higher luminosity.

\begin{figure*}
\includegraphics[scale=1.05, trim=1cm 7cm 0 0]{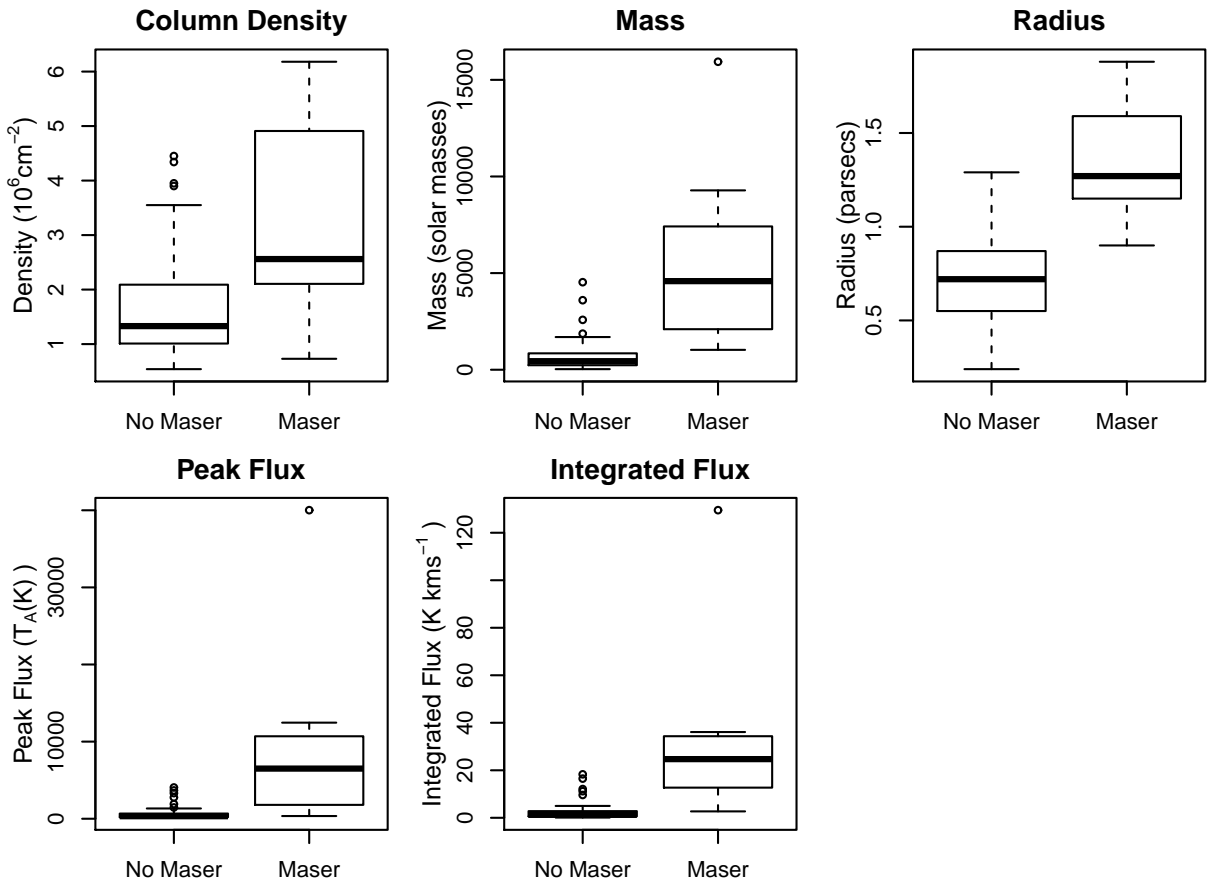}
\caption{Boxplots comparing the variables from Data Set 1 between the sources with an associated water maser and those without.  The outline in each of the boxplots represents the range between the first and third quartiles, with the median being the solid line horizontally through the box.  The vertical lines outside the box extend to the minimum and maximum values, with any outliers (values separated from the quartiles by more than one and a half times the interquartile range) shown separately as dots.  In this case, due to the very small number of samples being associated with a maser in this data set, the individual sample points are also plotted.}
\label{fig:BP-D} 
\end{figure*}

There are clear differences in the values of all the predictor variables between those sources with an associated water maser and those without, as is demonstrated by the boxplots shown in Figure~\ref{fig:BP-D}.  However, it should be noted that there are varying degrees of overlap in the ranges observed for the maser associated sources and those which are not.  The obvious difference in the distributions for all the predictor variables means that we might expect that they should all contribute to the classification and that the relative importance might also be similar.  The variable importance ratings returned by the random forest classification are a measure of the degree to which the classification trees utilised each predictor variable.  The five predictor variables available as inputs for the classification process were : peak flux density, source radius, total integrated flux density, dust mass (calculated assuming a temperature of 40 K and optically thin dust emission) and column density.  Using only source radius and the total integrated flux density provided the highest accuracy for random forests, logistic regression and LDA, while LDA using the ``normalised'' data (transformed using a log function, see Sec.~\ref{sec:LDA}) was able to utilise the column density too.  Table~\ref{tab:dust-var} shows the comparison of which of the predictor variables were included in the models based on their contributions to an increase in classification accuracy.  \citet{Breen+07} showed that their sample of water masers preferred denser, more massive and more luminous sources.  Our models indicated that the radius, luminosity and in the case of LDA on the normalised data, the density were important variables in predicting whether the sources were associated with a maser or not.  Our results are in agreement with \citet{Breen+07}, except that our models were not improved by inclusion of mass as a predictor variable.

\begin{table}
\caption{The predictor variables that increased the classification accuracy of the various methods for Data Set 1.  Random forests provides an internal calculation of the Mean Decrease in Accuracy (the higher the value, the more important the variable), logistic regression provides P-values (the lower the value, the more significant the variable's contribution to the model), and LDA provides no internal measurement of the importance of each variable, so it is just noted which variables were used (see Sec~\ref{sec:var}).}
\begin{tabular}{lrrcc} \hline
      & \multicolumn{1}{c}{\bf Random} & \multicolumn{1}{c}{\bf Logistic } & & \multicolumn{1}{c}{\bf Norm.} \\
      & \multicolumn{1}{c}{\bf Forests} & \multicolumn{1}{c}{\bf Reg.} & \multicolumn{1}{c}{\bf LDA} & \multicolumn{1}{c}{\bf LDA} \\

      \hline \hline
Radius & 10.68 & 0.1388 & Y & Y  \\
Int. Flux & 16.35 & 0.2485 & Y & Y \\
Density &       &       &       & Y \\
\hline
\end{tabular}
\label{tab:dust-var}
\end{table}

Table~\ref{tab:dust} summarises the results we obtained from cross validation of the three different classification techniques under consideration (for details see Sec.~\ref{sec:CV}).  Specificity values were high, due to the fact that the majority of the sources were not associated with masers, with the sensitivity values being lower in each case.  For Data Set 1, random forests  performed the best considering both sensitivity and specificity.  Notably, there are very few false positive classifications over all the models, which is most likely due to the data being unbalanced in that the majority of the samples were not associated with masers.  Another clear result is that performing LDA on the log transformed data increases the model's sensitivity, making it comparable to logistic regression in this case.  The advantage of transforming the data is also obvious in the ROC shown below in Figure~\ref{fig:ROC-D} (for an explanation on ROC curves, see Sec.~\ref{sec:ROC}).

\begin{table}
\caption{The results of cross-validating random forests, logistic regression and LDA  (without and with transformation of the predictor variables) classification and prediction for Data Set 1.}
\begin{tabular}{lrrrr} \hline
      & \multicolumn{1}{c}{\bf Random} & \multicolumn{1}{c}{\bf Logistic } & & \multicolumn{1}{c}{\bf Norm.} \\
      & \multicolumn{1}{c}{\bf Forests} & \multicolumn{1}{c}{\bf Reg.} & \multicolumn{1}{c}{\bf LDA} & \multicolumn{1}{c}{\bf LDA} \\

      \hline \hline
  True Neg. & 66 & 65 & 66 & 65  \\
  False Pos.&  0 &  1 &  0 &  1 \\
  False Neg.&  2 &  2 &  3 &  2  \\
  True Pos. &  5 &  5 &  4 &  5 \\
  Specificity\% & 100&98.5& 100&98.5 \\
  Sensitivity\% & 100&71.4&57.1&71.4 \\ \hline
   \end{tabular}

   \label{tab:dust}
\end{table}

\begin{figure}
\includegraphics[scale=0.8]{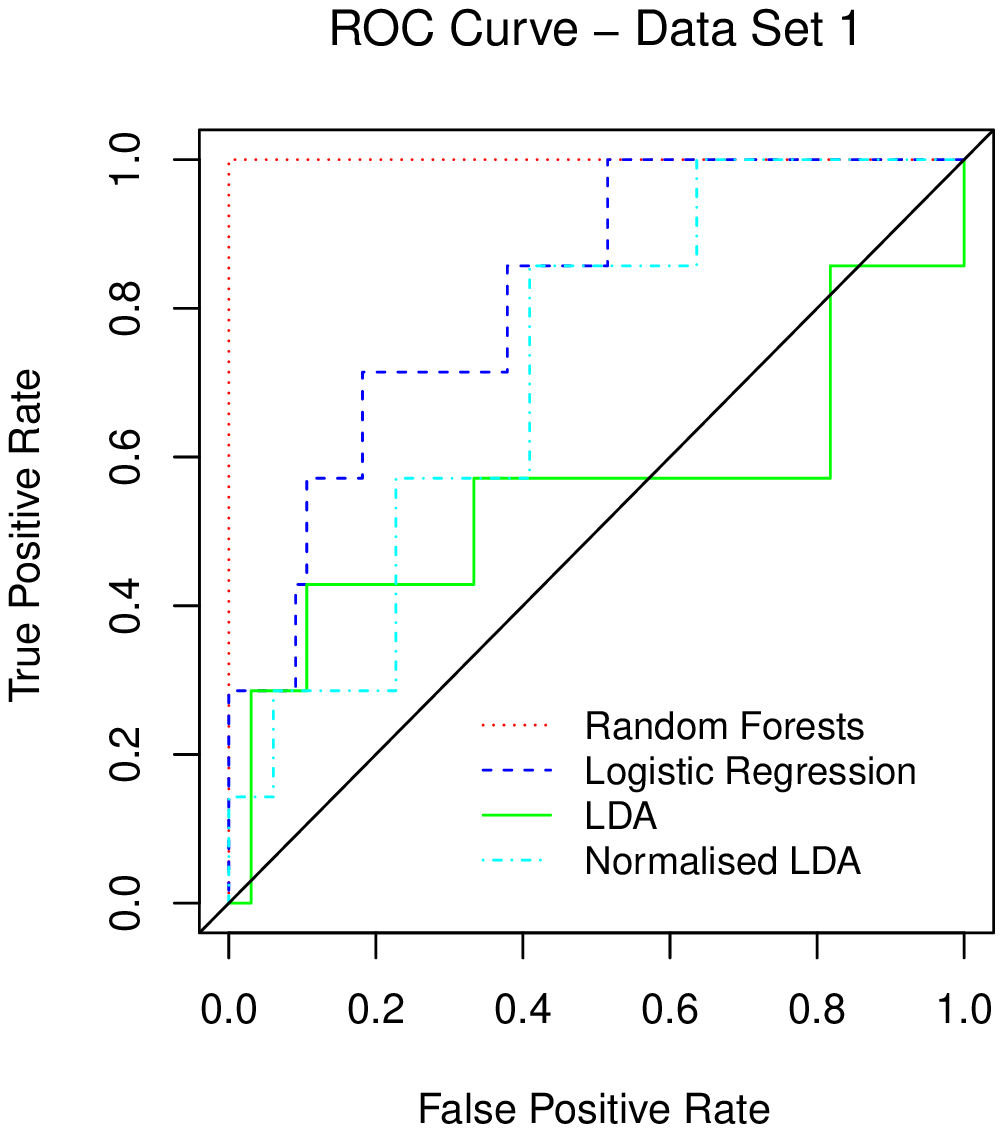}
\caption{Receiver operating characteristic curves showing the results of the cross validation for Data Set 1.  The diagonal line $y = x$ represents randomly classifying the samples, with half predicted as positive and half as negative.  For definitions of classification results, see Sec.~\ref{sec:classification}.}
\label{fig:ROC-D}
\end{figure}

Figure~\ref{fig:ROC-D} shows that LDA under-performs for Data Set 1, however when LDA is applied to the transformed data it is more accurate than logistic regression.  The relatively small data set causes the apparent steps in the plot and this is also evident for Data Set 2 in Figure~\ref{fig:ROC-S}.  The ROC curves for Data Set 3 (Fig.~\ref{fig:ROC-B}) are much smoother because there are 214 samples rather than 73, or 32.  Despite the apparent steps in the ROC curves, the plot very clearly shows the most accurate classification technique for this data set (the non-parametric method of random forests) and the least accurate (the parametric method of LDA using untransformed data).

\begin{figure}
\includegraphics[scale=0.8]{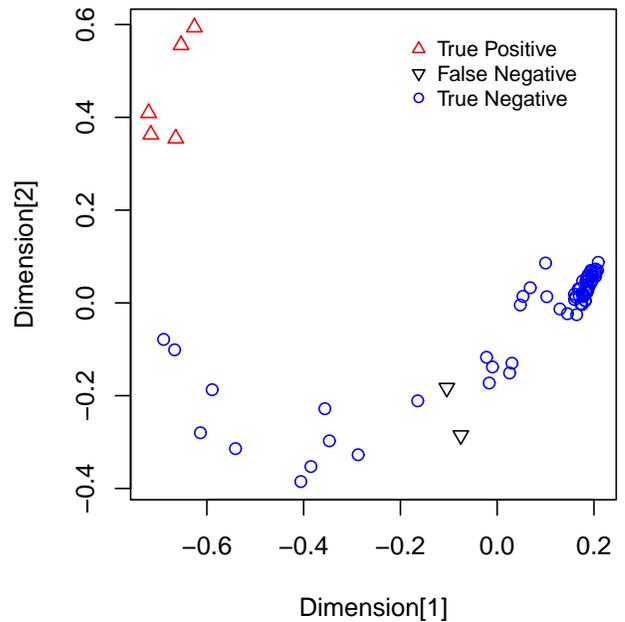}
\caption{MDS plot of the proximity values produced by the random forest classification (Data Set 1).  The values on the axes are arbitrary, the graph just compares relative magnitudes.  The closer two points are on the plot, the more similar their properties as determined by the random forest classification.  ``Border-line" classifications were samples with a predicted maser association between 45 and 55\%.}
\label{fig:MDS-D}
\end{figure}

Figure~\ref{fig:MDS-D} shows a multi-dimensional scaling (MDS) plot for the full data set. MDS plots give a visual representation of the distances between proximities identified in the random forest implementation; sources that the random forest process identifies as being similar are clustered within the MDS plot.  The distance values are arbitrary, they are simply relative magnitudes, plotted here as Dimension[1] and Dimension[2].  Figure~\ref{fig:MDS-D} shows the four correctly identified maser sources in a group at the top-left, separated from the non-maser sources.  ``Border-line" classifications were samples with a predicted maser association between 45 and 55\%, with the last correctly classified maser shown just below the others as such.  The model was not sensitive enough to detect the differences in the predictor variables for the other maser-associated sources (which is why they were classified as not having an associated water maser).  This is probably due to the small number of sources in this data set. 

\subsection{The properties of water maser-associated YSOs in the LMC}
\label{sec:SED}

\citet{Gruendl+09} used the {\em Spitzer Space Telescope} Surveying Agents of Galaxy Evolution (SAGE) Legacy programme data \citep{Meixner+06}, along with other public data sets to identify high- and intermediate-mass young stellar objects (YSOs) in the Large Magellanic Cloud (LMC).  \citeauthor{Gruendl+09} identified 855 definite YSOs in the LMC and compiled near- and mid-infrared photometric measurements for the sample.  \citet{Ellingsen+10} made Australia Telescope Compact Array (ATCA) observations for the 22~GHz transition of water towards all known star formation maser sites in the LMC, resulting in a total of 13 water masers in the LMC for which positions are known to arcsecond accuracy.  The fields observed for the water maser observations included a total of 32 sources from the \citet{Gruendl+09} YSO catalogue.  Of the 13 water masers, 11 are within 2 arcseconds of a \citeauthor{Gruendl+09} YSO, meaning that from a total catalogue of 855 sources there are 11 which are known to have an associated water maser and 22 which are known not to. The 33 sources for which there is information on whether or not they have an associated water maser can be used as a training set for classification/prediction.  

\citet{Ellingsen+10} used the infrared data from \citet{Gruendl+09} to construct the spectral energy distribution (SED) of each of the YSOs using the online SED-fitter of \citet{Robitaille+07} and this forms Data Set 2.  For some wavelength ranges the infrared data for the \citet{Gruendl+09} sample is incomplete, hence there is missing data.  However, the results of the SED modelling contain no missing data (although there is likely to be greater uncertainty in the fitted SED parameters for those sources which have less infrared photometric measurements contributing to the fitting process).  All available information about a source is incorporated into the SED model.  According to \citet{Ellingsen+10}, there is very little variation in the amount of information available for each SED fit, with between seven and nine infrared intensities available for each source and in the majority of cases the chi-squared values for the resulting SED fits are reasonable.  Due to the large number of sources modelled, we made no attempt to remove the sources where this was not the case, with the exception of one maser-associated source with more missing data than the others (making our training sample 32 with 10 known masers, and the total data set 854).

\begin{figure*}
\includegraphics[scale=1.05, trim=1cm 7cm 0 0]{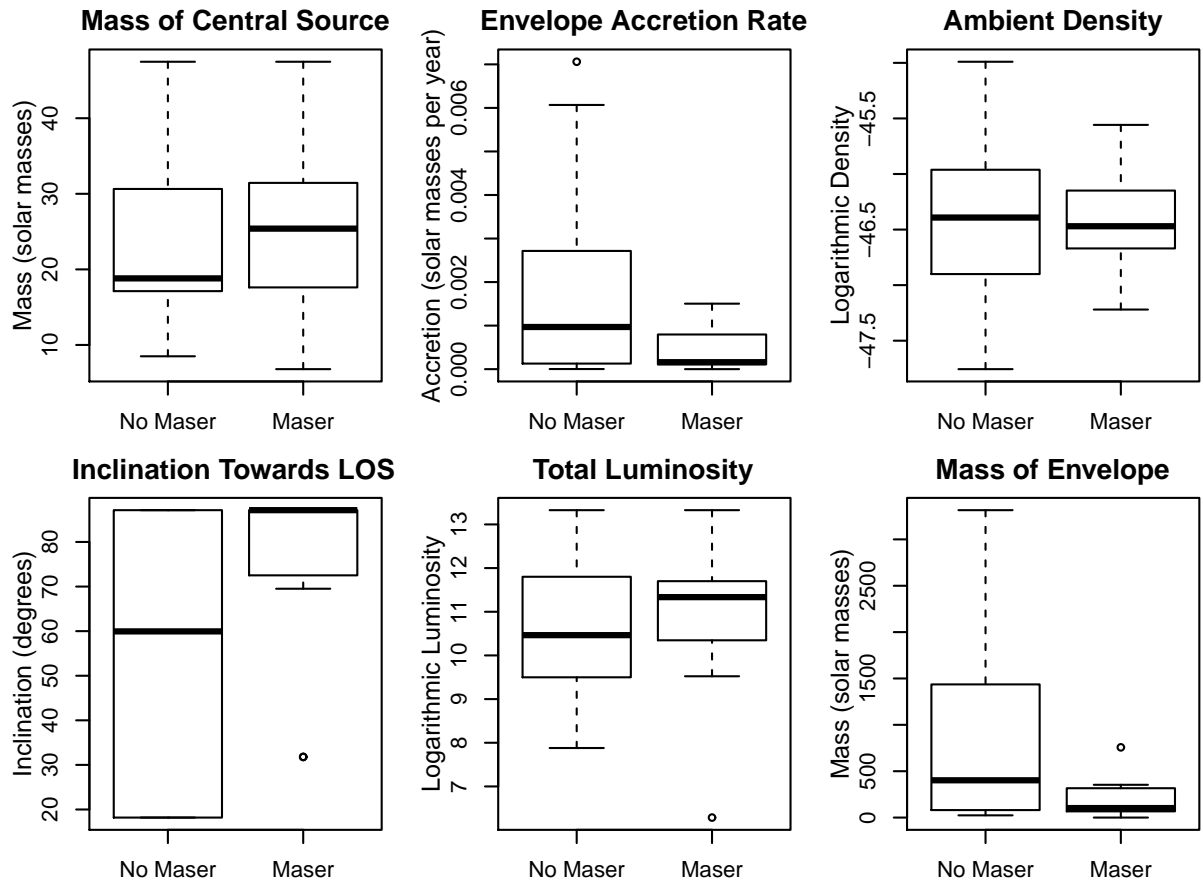}
\caption{Only the predictor variables from Data Set 2 showing noticeable differences between those YSO with and without an associated water maser are shown.  Due to the very small data set, the individual sample points are also plotted.  Some of the variables are on logarithmic scales to better illustrate the differences.  For an explanation of boxplots, see Fig.~\ref{fig:BP-D}. }
\label{fig:BP-S}
\end{figure*}

Fifteen predictor variables were extracted from the SED fitting results; distance to the source, age, radius, mass and temperature of the central source, envelope accretion or infall rate, outer and inner radius of the envelope, cavity opening angle, disc mass, ambient density, inclination of source to line of sight (LoS), average integrated flux density (from the outside of the YSO to the stellar surface, along the LoS), total luminosity, and mass of the envelope.  Table~\ref{tab:sed-var} shows which variables were used in each model and how they contributed to that model.  Across the different methods, the most important predictor variables appeared to be the mass of the central source, the outer envelope radius, the inclination towards the LoS, and the mass of the envelope.  In comparison, \citet{Ellingsen+10} found that the majority of YSOs with an associated water maser have high luminosities, central masses and ambient densities.  They also tend to have redder infrared colours than those YSOs which are not associated with a maser.  The distributions of the high-importance variables are shown in Figure~\ref{fig:BP-S}.

Unlike Data Set 1, these data include some sources where the maser association is known (32) and some where it is unknown (822).  This means predictions can be made on the unknown sources.  To test how well the various methods will generalise to data where maser association is unknown, a cross-validation was applied.  For a full description of the technique used, see Section~\ref{sec:CV}.  The predictions were then compared with the actual maser association.  The results are given in Table~\ref{tab:SED-CV}.

\begin{table}
\caption{The predictor variables that increased the classification accuracy of the various methods for Data Set 2.  The value given for random forests is the Mean Decrease in Accuracy, while logistic regression provides P-values.  The most important variables in logistic regression and random forest models are shown in bold.  For further explanation see Table~\ref{tab:dust-var}.}
\begin{tabular}{lrrcc} \hline
      & \multicolumn{1}{c}{\bf Random} & \multicolumn{1}{c}{\bf Logistic } & & \multicolumn{1}{c}{\bf Norm.} \\
      & \multicolumn{1}{c}{\bf Forests} & \multicolumn{1}{c}{\bf Reg.} & \multicolumn{1}{c}{\bf LDA} & \multicolumn{1}{c}{\bf LDA} \\

      \hline \hline
Distance &       &       &       & Y \\
Age   &       &       & Y &  \\
Mass  & \bf{3.943} &  \bf{0.0805} & Y &  \\
Radius & 2.193 &  0.395 & Y & Y \\
Temperature & 2.091 &       & Y &  \\
Accretion & 2.257 &       & Y & Y \\
Outer Env. &       & 0.1609 & Y &  \\
Inner Env. & 1.099 &       & Y & Y \\
Cavity Angle & \bf{3.891} &       & Y & Y \\
Disc Mass &       &       & Y & Y \\
Amb. Density &       &  0.1935  &       &  \\
Inclination & 2.677 &       & Y & Y \\
Av. Int. Flux &       &       & Y &  \\
Total Lum. & 1.047 &       &       &  \\
Env. Mass & \bf{4.202} & 0.2765 & Y & Y \\
\hline
\end{tabular}
\label{tab:sed-var}
\end{table}

\begin{figure}
\includegraphics[scale=0.8]{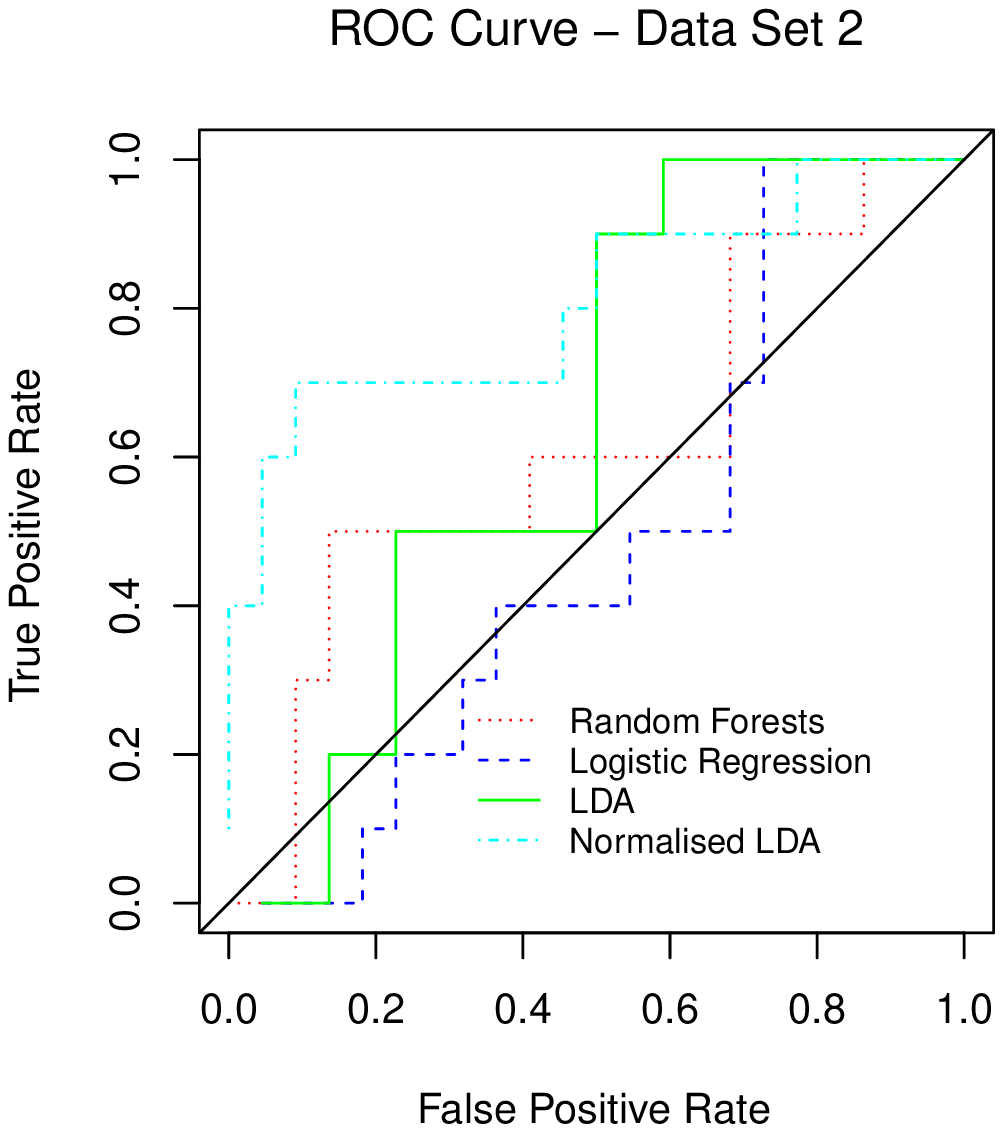}
\caption{Receiver operating characteristic curves showing the results of the cross validation for Data Set 2.  The diagonal line $y = x$ represents randomly classifying the samples, with half predicted as positive and half as negative.  For a full description of a ROC curve, see Section~\ref{sec:ROC}.}
\label{fig:ROC-S}
\end{figure}

The SED Data Set 2 had a fairly small number of entries with known maser status (32), but a large number of possible predictor variables (15).  The results for the cross validation are shown in Table~\ref{tab:SED-CV} and in the form of ROC curves in Figure~\ref{fig:ROC-S}. 

The cross validation results show that overall, the sensitivity values were quite poor, with LDA performed on the normalised data being the most accurate method.  This was possibly due to the small data set, methods could not construct an accurate model using only 29 sources and then predicting on the remaining 3.  These results can be visualised in a ROC curve shown in Figure~\ref{fig:ROC-S}, where in a number of cases the models fall below the $y=x$ diagonal line, meaning that they perform \textit{worse} that simple random classification with a 50\% chance of a source being associated with a maser.  Due to these poor results, we decided not to use this training data to make predictions on the remaining 822 sources with unknown maser status.

\begin{figure}
\includegraphics[scale=0.8]{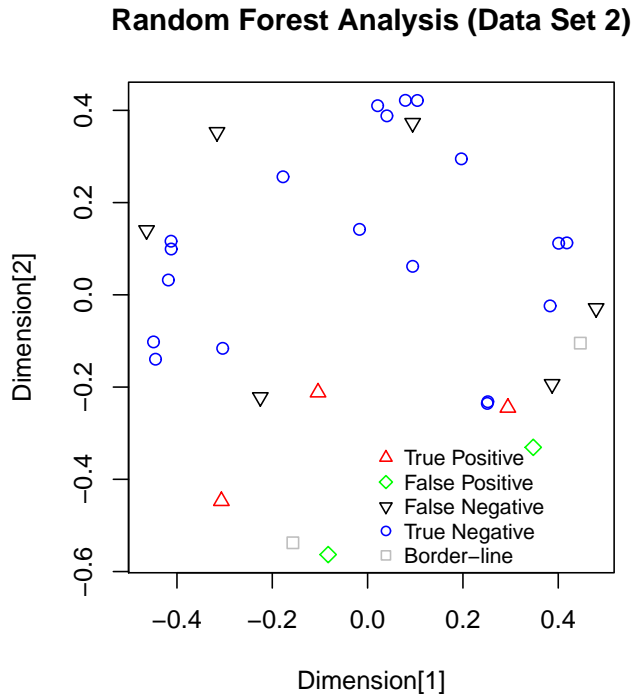}
\caption{The MDS plot for the random forest model used to predict potential YSOs with an associated water maser in the LMC (Data Set 2).  ``Border-line" predictions were samples with a predicted maser association between 45 and 55\%.  For details on multidimensional plots in random forest analysis, see Fig.~\ref{fig:MDS-D}.}
\label{fig:MDS-S}
\end{figure}

It is evident from the MDS plot in Figure~\ref{fig:MDS-S} why the random forest model had a low sensitivity; the data is not clustered in groups to the same extent as Data Set 1 (Fig.~\ref{fig:MDS-D}).  The known maser sources are represented by the black and red triangles, while the blue circles and green diamonds represent the known non-maser sources.  This could be due to the model's inability to condense the 15 predictor variables (equivalent to 15 dimensions) into a two-dimensional plot, or another bi-product of the small sample size.  

\begin{table}
\caption{The results of cross-validating random forests, logistic regression and LDA classification and prediction for Data Set 2 (association of water masers with infrared YSO in the LMC) using the full sample of 32 sources with known water maser association status as the training sample.  For definitions of classification results, see Sec.~\ref{sec:classification}.}
\begin{tabular}{lrrrr} \hline
      & \multicolumn{1}{c}{\bf Random} & \multicolumn{1}{c}{\bf Logistic } & & \multicolumn{1}{c}{\bf Norm.} \\
      & \multicolumn{1}{c}{\bf Forests} & \multicolumn{1}{c}{\bf Reg.} & \multicolumn{1}{c}{\bf LDA} & \multicolumn{1}{c}{\bf LDA} \\

      \hline \hline
  True Neg.   & 20 & 17 & 15 &  20 \\
  False Pos.  &  2 &  5 &  7 &   2 \\
  False Neg.  &  6 &  8 &  5 &   4 \\
  True Pos.	  &  4 &  2 &  5 &   6 \\
  Specificity\%	  &90.9&77.3&68.2&90.9 \\
  Sensitivity\%	  &40.0&20.0&50.0&60.0 \\ \hline
   \end{tabular}
   \label{tab:SED-CV}
\end{table}

In summary, the variables that had the most influence over the various classification models were mass of the central star, the outer envelope radius, the inclination towards the LoS, and the mass of the envelope (see Table~\ref{tab:sed-var}).  The likelihood of a YSO being associated with a water maser source did not appear to depend heavily on variables such as the age of the source, mass of the disc, ambient density or average integrated flux.  \citet{Ellingsen+10} applied Mann-Whitney tests to the different variables to find the difference in the medians of the distributions of those associated with masers and those not associated.  Statistically significant differences were found in the data for the mass of the central star, the outer radius of the envelope, ambient density, inclination towards the line of sight and the total luminosity; results that agree with our analysis.  It was previously suggested that the inclination angle is one of the most influential predictors in determining the SED for young stellar objects \citep{Robitaille+06}.  As a result of the orientation of the cavity, the inclination angle dictates the contribution from the inner, hotter regions of the envelope to the SED.  Hence, our classification results here agree with those from previous studies, indicating that the physical variables mentioned above are likely to dictate water maser-association with certain YSOs in the LMC.

\subsection{The properties of dust continuum emission associated with class~I methanol masers} \label{sec:classI}

The final data set we investigated (hereafter, Data Set 3) was a search for 95~GHz class~I methanol masers targeted towards regions selected on the basis of both their emission at mid-infrared and millimetre wavelength ranges \citep{Chen+12}.  The mid-infrared data was taken from the {\em Spitzer Space Telescope} GLIMPSE (Galactic Legacy Infrared Mid-Plane Survey Extraordinaire) program, which provides photometric measurements in four wavelength bands \citep[3.6, 4.5, 5.8 and 8.0 $\mu$m;][]{Benjamin+03,Churchwell+09}, while the millimetre continuum data was from the Bolocam Galactic Place Survey BGPS \citep{Aguirre+11}.  The motivation for this survey was a previous search for 95~GHz class~I methanol masers by \citet{Chen+11}. The authors targeted infrared sources for which GLIMPSE images show extended emission with an excess in the 4.5-$\mu$m band (thought to indicate an outflow from a high-mass YSO).  It was found that those GLIMPSE sources with an associated BGPS source (54 of the 62 sources which lay within the BGPS region) were much more likely to exhibit class~I methanol maser emission.  \citet{Chen+11} also found that the GLIMPSE sources with redder mid-infrared colours were more likely to be associated with methanol masers and the higher the mass and density of the BGPS dust clump, the stronger the class I maser emission.

\citet{Chen+12} used the results of \citet{Chen+11} to identify 420 sources detected in both the {\em Spitzer} GLIMPSE and BGPS catalogues as likely to have an associated class~I methanol maser.  They then observed a random selection of 214 of these sources and detected 95~GHz class~I methanol masers towards 62 (hence 152 non-detections).  For the classification process we used only the data from the BGPS catalogue (version 1.0) which contains a total of 8358 sources \citep{Aguirre+11}. The predictor variables used in the classification models for Data Set 3 were the angular size of the major and minor axis of the dust clump, as well as its position angle, deconvolved angular radius and 1.1 mm flux density within apertures of diameter 40, 80 and 120 arcseconds and the integrated flux density.  As with the classification of the other two data sets, here each of the models were optimised by omitting superfluous variables, as well as those that decreased the models' accuracy.  Both logistic regression and random forests utilised all eight variables, while LDA performed better without including all of them (see Table~\ref{tab:bol-var}).

\begin{figure*}
\includegraphics[scale=1.05, trim=1cm 1cm 0 0]{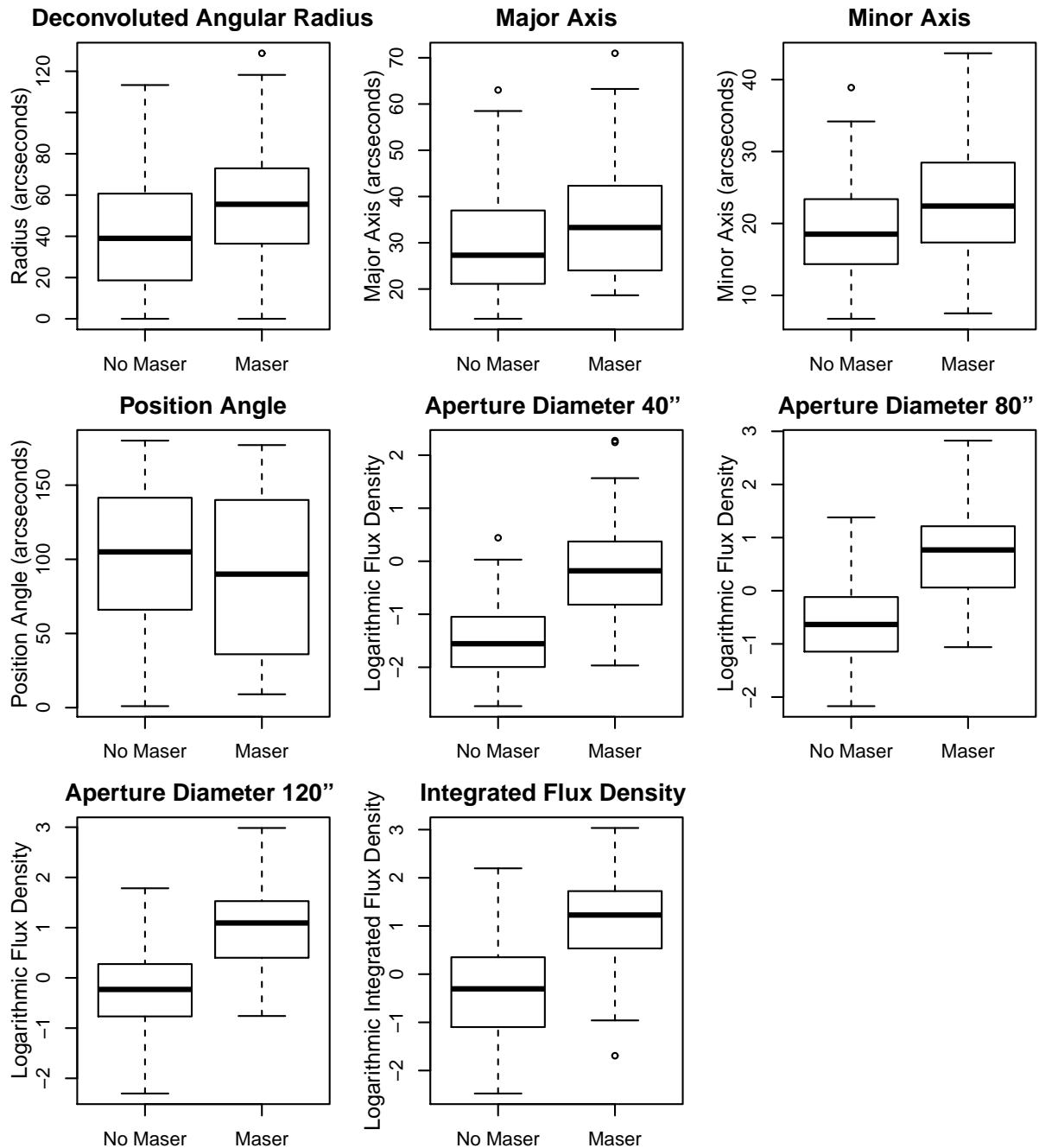}
\caption{The variables used in the classification and prediction of Data Set 3.  Some of the variables are on logarithmic scales to better illustrate the differences.  For an explanation of boxplots, see Fig.~\ref{fig:BP-D}.}
\label{fig:BP-B}
\end{figure*}

This data set was the primary focus of our analysis, as it has a training set with several hundred sources, including a large number of detections and there are also a large number of BGPS sources which have not been searched for class I methanol maser emission (8144) which provide the opportunity to make testable predictions. 

\begin{table}
\caption{The predictor variables that increased the classification accuracy of the various methods for Data Set 3.  The value given for random forests is the Mean Decrease in Accuracy, while logistic regression provides P-values.  The most important variables in logistic regression and random forest models are shown in bold.  For further explanation see Table~\ref{tab:dust-var}.}
\begin{tabular}{lrrcc} \hline
      & \multicolumn{1}{c}{\bf Random} & \multicolumn{1}{c}{\bf Logistic } & & \multicolumn{1}{c}{\bf Norm.} \\
      & \multicolumn{1}{c}{\bf Forests} & \multicolumn{1}{c}{\bf Reg.} & \multicolumn{1}{c}{\bf LDA} & \multicolumn{1}{c}{\bf LDA} \\

      \hline \hline
Major Axis & 4.124 & 0.2781 & Y & Y \\
Minor Axis & 10.28 & 0.3867 & Y & Y \\
Position Angle & 1.758 & \bf{0.0952} &       & Y \\
Angular Radius & 9.218 & 0.3252 &       &  \\
40 arcseconds & \bf{27.94} & \bf{0.0312} & Y & Y \\
80 arcseconds & \bf{21.81} & 0.4267 & Y &  \\
120 arcseconds & 14.24 & 0.6531 & Y &  \\
Int. Flux Den. & 15.31 & 0.1158 & Y & Y \\
\hline
\end{tabular}
\label{tab:bol-var}
\end{table}

The variables with high importance in the random forests calculations were the flux densities (each of the 40, 80 and 120 arcsecond aperture values and the integrated) and also the angular size of the minor axis.  The most important variable was the flux density within 40 arcseconds (the smallest angular scale measured by Bolocam).  Logistic regression also found the flux density within 40 arcseconds to be the most important variable with a P-value of 0.0312, with the next most significant variable being the position angle with a P-value of 0.0952, while the 80 arcsecond flux had the next highest contribution.  This is consistent with the results of \citet{Chen+12} which showed that class~I masers were preferentially associated with sources with the highest beam averaged column density (which is directly proportional to the 40 arcsecond flux density).  There is no physical reason why the position angle of the dust clump would effect the likelihood of a dust clump having an associated class I methanol maser, but when this variable was omitted from the classification, the accuracy of the models decreased.  However, while the P-value suggests that the position angle is a significant predictor variable in logistic regression, the change in accuracy was not significant compared to that of the other variables.  This suggests that we can dismiss it as an artefact of the classification method, but it does serve as a reminder to view results such as this with a degree of scepticism.  It is also worth noting that random forests presented it with the lowest variable importance.

 \begin{table}
 \caption{The results of cross-validating random forests, logistic regression and LDA classification and prediction for Data Set 3 (class I methanol masers associated with {\em GLIMPSE} sources).  Fig.~\ref{fig:ROC-B} shows the ROC curve for each of the models.  For definitions of classification results, see Sec.~\ref{sec:classification}.}
 \begin{tabular}{lrrrr} \hline
       & \multicolumn{1}{c}{\bf Random} & \multicolumn{1}{c}{\bf Logistic } & & \multicolumn{1}{c}{\bf Norm.} \\
       & \multicolumn{1}{c}{\bf Forests} & \multicolumn{1}{c}{\bf Reg.} & \multicolumn{1}{c}{\bf LDA} & \multicolumn{1}{c}{\bf LDA} \\
 
       \hline \hline
   True Neg.  & 141 & 142 & 148 & 145 \\
   False Pos. &  11 &  10 &   4 &   7 \\
   False Neg. &  21 &  24 &  31 &  23 \\
   True Pos.  &  41 &  38 &  31 &  39 \\
   Specificity\%  & 92.8& 93.4& 97.4&95.4 \\
   Sensitivity\%  & 66.1& 61.3& 50.0&62.9 \\ \hline
    \end{tabular}
    \label{tab:Bol-CV}
 \end{table}

\begin{figure}
\includegraphics[scale=0.8]{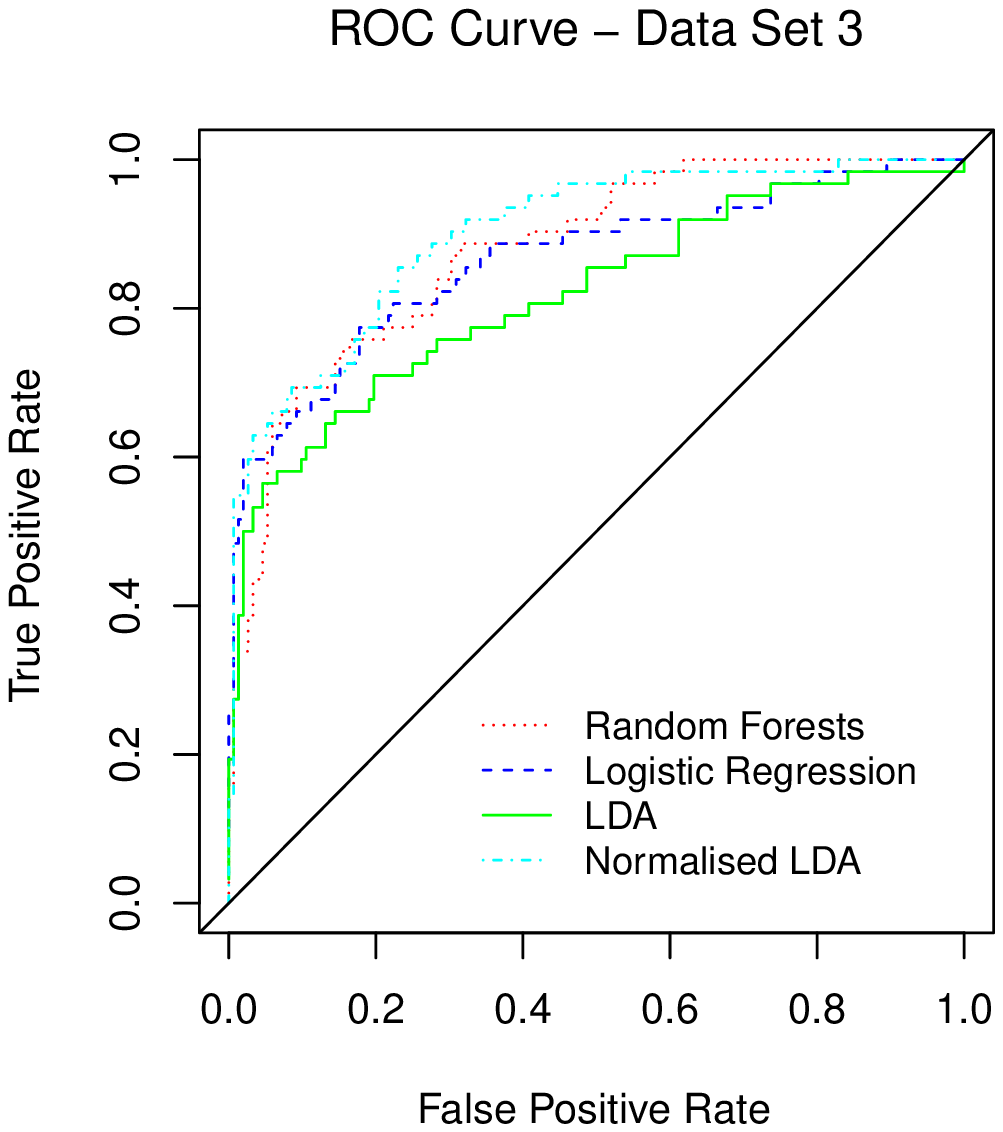}
\caption{Receiver operating characteristic curves showing the results of the cross validation for Data Set 3.  The diagonal line $y = x$ represents randomly classifying the samples, with half predicted as positive and half as negative. }
\label{fig:ROC-B}
\end{figure}

Table~\ref{tab:Bol-CV} shows the results of the cross-validation of the different classification techniques used on Data Set 3 (see Sec.~\ref{sec:CV}).  Here, random forests offered the highest sensitivity, while surprisingly performing LDA on the untransformed data produced the highest specificity.  This is the first instance in our studies where transforming the data set to be closer to a normal distribution decreased the performance of LDA,  although the decrease was minor (2\%) and likely not significant.  The ROC curve in Figure~\ref{fig:ROC-B} gives a more complete representation of the models' capabilities, showing that random forests, logistic regression and LDA using the transformed data performed to similar standards, while generally LDA  on the untransformed data performed the worst.  

\begin{figure*}
\includegraphics[scale=0.9]{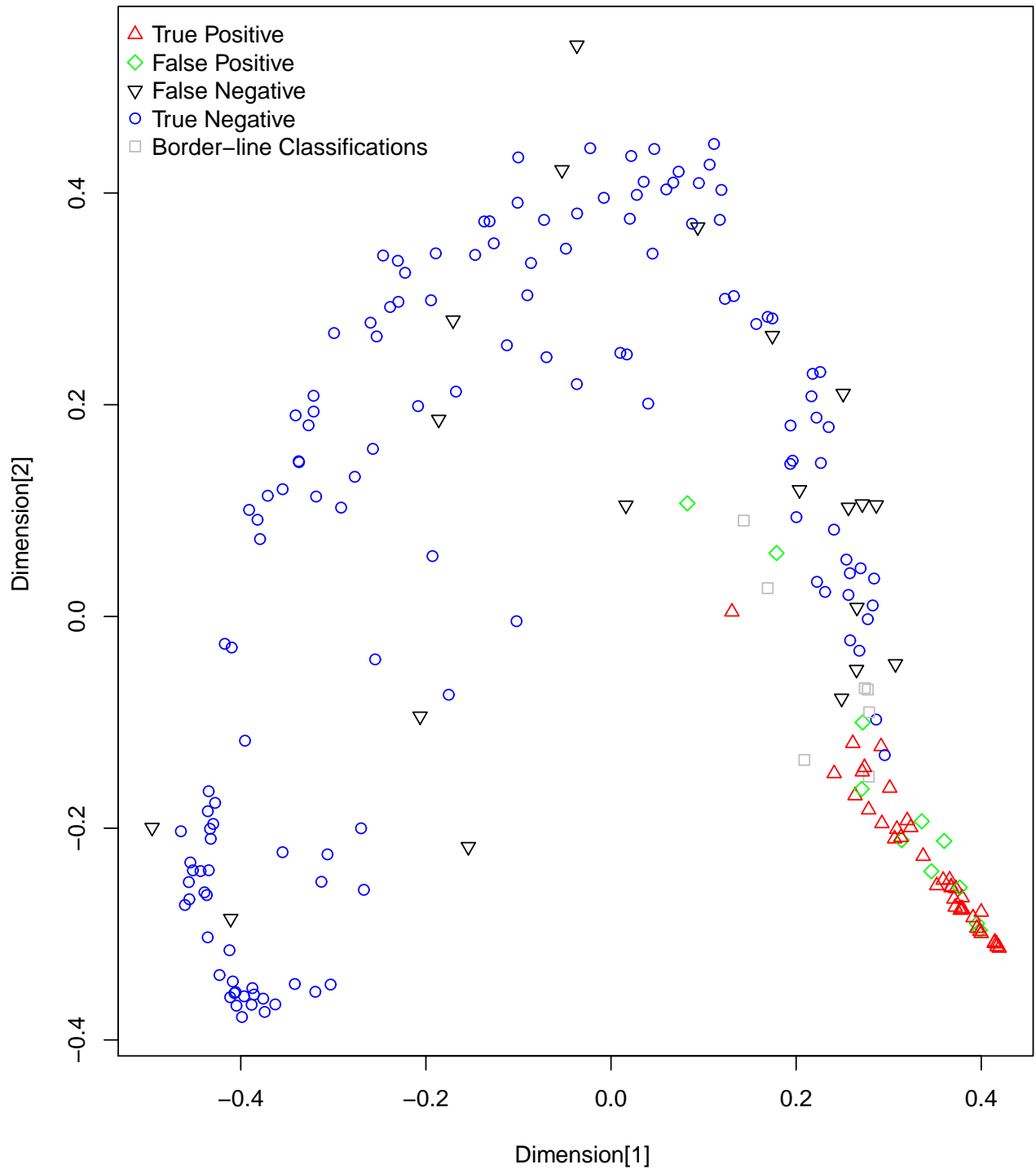}
\caption{The MDS plot for the random forest model used to predict potential millimetre dust-clumps with associated class~I methanol masers (Data Set 3).  ``Border-line" classifications were samples with a predicted maser association between 45 and 55\%.  For details on multidimensional plots in random forest analysis, see Fig.~\ref{fig:MDS-D}.}
\label{fig:MDS-B}
\end{figure*}

Figure~\ref{fig:MDS-B} shows the MDS plot for the random forest model generated using all the training data for Data Set 3.  It is clear that the maser-associated dust clumps are generally located in the bottom-right region of the plot.  Comparing this plot to the MDS plots for the other two Data Sets (Fig.s~\ref{fig:MDS-D} \& \ref{fig:MDS-S}), we can see that the maser associated sources are more clearly separated from those without a maser-association.  The green squares in Figure~\ref{fig:MDS-B} represent sources which the random forests model predicts to have an associated class~I methanol maser, but for which the observations of \citet{Chen+12} did not detect a maser.  Many of these sources lie very close on the MDS plot to others where a maser was detected and it may be that some of these non-detections have a weak class~I methanol maser which was not detected by \citeauthor{Chen+12} due to the limited sensitivity of those observations.  The 95~GHz class~I methanol masers are in the same transition family as the best studied class~I methanol maser transition at 44~GHz.  In general the 44~GHz class~I methanol masers have a peak flux density approximately a factor of 3 greater than the 95~GHz maser emission in the same source \citep{Valtts+00}.  These sources would be good candidates for sensitive observations in the 44~GHz transition to more robustly determine if they are associated with class~I methanol masers.

The classification models we have developed can also be used to predict which of the BGPS sources that were not observed by \citet{Chen+12} are the best candidates for having an associated class~I methanol maser.  Since we have four different classification models we can compare the results of each, as those sources identified by all, or most of the models would be expected to be the promising targets for further searches. 

 \begin{table}
 \caption{The classification results on the training data subset (where the maser presence is known), and the number of predicted masers from the 8144 sources for which maser presence is unknown, using Data Set 3 (class I methanol masers associated with {\em GLIMPSE} sources).  For definitions of classification results, see Sec.~\ref{sec:classification}.}
 \begin{tabular}{lrrrr} \hline
       & \multicolumn{1}{c}{\bf Random} & \multicolumn{1}{c}{\bf Logistic } & & \multicolumn{1}{c}{\bf Norm.} \\
       & \multicolumn{1}{c}{\bf Forests} & \multicolumn{1}{c}{\bf Reg.} & \multicolumn{1}{c}{\bf LDA} & \multicolumn{1}{c}{\bf LDA} \\
 
       \hline \hline
   True Neg.  & 140 & 145 & 149 & 147 \\
   False Pos. &  12 &   7 &   3 &   8 \\
   False Neg. &  21 &  22 &  30 &  22 \\
   True Pos.  &  41 &  40 &  32 &  40 \\
   Specificity\%  &92.1 &95.4 &98.0 &96.7 \\
   Sensitivity\%  &66.1 &64.5 &51.6 &64.5 \\ \hline
   Predictions & 632 & 405 & 334 & 460 \\ \hline
    \end{tabular}
    \label{tab:Bol-pred}
 \end{table}

For the prediction model, as with Data Set 2, we grew a random forest using 3000 trees (instead of the default 500, see Sec.~\ref{sec:SED}).  Table~\ref{tab:pred} in the Appendix lists the 739 BGPS sources which were predicted to have an associated class~I methanol maser by one or more of the four classification models for the 8144 BGPS sources which have not yet been searched.  Table~\ref{tab:Bol-pred} shows that of the 8144 potential BGPS target sources random forests predicts 632 to have an associated class~I methanol maser and this is significantly more than any of the other classification models.  There are 242 of the 8144 BGPS sources which all models predict will have an associated class~I methanol maser and these will be the prime targets for future searches.  Table~\ref{tab:pred_2} shows the number of sources predicted to be masers by one or two classification methods, which should be considered if there is sufficient time to search additional targets.

\begin{table}
\caption{Number of maser predictions on sources from Data Set 3 shared by two classification methods, with 242 sources predicted to be masers using all four methods.}
\label{tab:pred_2} 
 \begin{tabular}{l|cccc} \hline
       & \multicolumn{1}{c}{\bf Random} & \multicolumn{1}{c}{\bf Logistic } & & \multicolumn{1}{c}{\bf Norm.} \\
       & \multicolumn{1}{c}{\bf Forests} & \multicolumn{1}{c}{\bf Reg.} & \multicolumn{1}{c}{\bf LDA} & \multicolumn{1}{c}{\bf LDA} \\
       \hline
 {\bf Random} & \multirow{ 2}{*}{632} & \multirow{ 2}{*}{364} & \multirow{ 2}{*}{317} & \multirow{ 2}{*}{377} \\
 {\bf Forests} & & & & \\
 & & & & \\
    & & & & \\
 {\bf Logistic} &  & \multirow{ 2}{*}{405} & \multirow{ 2}{*}{254} & \multirow{ 2}{*}{371} \\
  {\bf Reg.} & & & & \\
   & & & & \\
   \multirow{ 2}{*}{{\bf LDA}} & & & \multirow{ 2}{*}{334} & \multirow{ 2}{*}{256} \\
   & & & & \\
   & & & & \\
  {\bf Norm.} & & & &  \multirow{ 2}{*}{460} \\
  {\bf LDA} & & & & \\    
     & & & & \\
       \hline
    \end{tabular}
 \end{table}

\section{Discussion} \label{sec:discussion}

We applied three different classification techniques to three different searches for interstellar masers to investigate each technique's performance.  We show the classification and prediction results of LDA performed on both the normally distributed data and the un-transformed data to demonstrate the difference (see Sec.~\ref{sec:LDA}).  In most cases, LDA performs significantly better when applied to transformed data. 

All three methods of classification (both parametric and non-parametric) used on Data Set 1 returned high values for both sensitivity (correctly classifying sources associated with masers) and specificity (correctly classifying non-maser sources).  The highest accuracy was achieved through the non-parametric method of random forests, which in this case classified every source correctly.  

Data Set 2 had a relatively small training sample (32 sources) compared with the number of predictor variables (15), and we found here that LDA appeared to give the best results, while random forests and logistic regression performed quite poorly in correctly identifying sources associated with a maser.

For Data Set 3, which has more than 50 detections and more than 150 non-detections, the non-parametric random forests had the highest sensitivity, while the parametric method of LDA performed on the untransformed data had the lowest, but also had the highest specificity.  Considering both sensitivity and specificity, logistic regression and random forests were the most accurate methods.  

Based on the predictions of \citet{Breiman+01a} our initial expectation was that given sufficient training data the complex relationship between the predictor variables and the presence or absence of a related astrophysical phenomenon would be more accurately represented by a non-parametric approach than a simple linear model.  However, the training data sets were relatively small, and so made it difficult for the models to capture and convey all the information contained within the predictor variables.  We find that random forests does perform relatively better for the largest data set, but in this case it is comparable with the accuracy of the non-parametric techniques, not superior to them.  It may be that in order to outperform parametric methods the non-parametric techniques require still larger amounts of training data.  However, it is more likely that for Data Set 3 all techniques approximately reach the limit of the information available within the measured parameters of the data.

There are a number of factors related specifically to the data which will lead to limitations in the accuracy of any  classification model developed using it.  One factor is the intrinsic measurement uncertainty for parameters such as the flux density, angular size etc., which can influence the results directly in the sense that it is always possible that given observations with greater sensitivity additional sources would be detected.  However, the absence of these weaker sources does more than simply qualifying the question that is being answered by the  classification model.  For example, the intensity of astrophysical masers depends in a complex and non-linear manner on the physical parameters of the environment and some of these parameters may not be represented either directly or indirectly in any of the predictor variables being used as inputs to the  classification methods.  A second, less obvious factor which may limit the accuracy of  classification techniques is that for derived parameters there are often implicit assumptions.  For example the calculation of the mass of the dust clumps used for Data Set 1 assumes that the emission at 1.2~mm wavelength is optically thin (likely a reasonable assumption), and that the dust is at a constant temperature of 40~K for all the dust.  This second assumption is necessary because we do not have any information on the specific temperature distribution of the dust, but it inevitably leads to systematic errors in the relative mass calculated for regions where the true dust temperature is on average higher (or lower) than the assumed value. Similarly, the distance to individual sources has been estimated using kinematic distance models, which on average provide a reasonable estimate, but which can lead to significant errors for individual sources.  It is also highly probable that our sensitivity values obtained after cross validation of the classification techniques were poor due to the unbalanced nature of the data, in that for all three data sets, the vast majority of the samples are not associated with masers.

\citet{Breen+11b} tested the binomial generalised linear model of \citep[][Data Set 1]{Breen+07} by searching for 22~GHz water masers towards 267 dust clumps. They found a high detection rate towards dust clumps for which the binomial GLM predicted a probability of greater than 10 \% for the presence of a water maser (20 of 27 sources). They also found that while the detection rate dropped for sources for which the model predicted a lower probability of having an associated water maser, a substantial fraction of water masers (approximately 70 \%) were detected towards sources for which the model predicted a probability of less than 1 \%.  \citet{Breen+11b} show that unreliable distance estimates for many of the dust clumps is in part responsible for the misclassification. This is consistent with the assertion we make above that the combination of measurement and systematic uncertainties in the underlying data ultimately limit the accuracy which can be obtained with any  classification technique.

\begin{figure}
\includegraphics[scale=0.4]{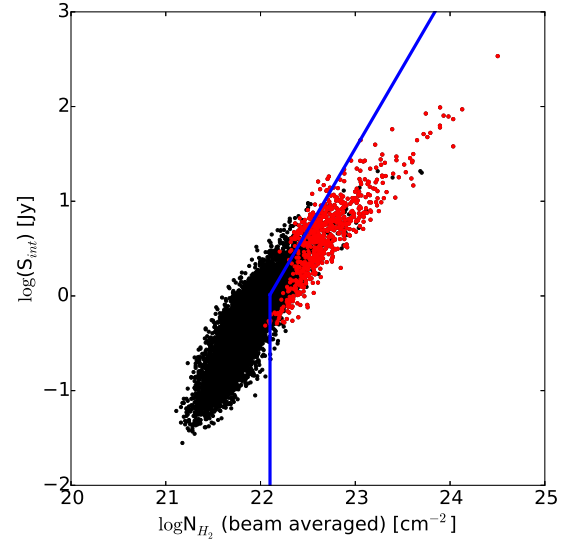}
\caption{The integrated flux density versus the beam averaged H$_2$ column density for the 8144 BGPS sources not searched for class~I methanol masers by \citet{Chen+12}.  Sources for which one or more of the  classification models predicts the presence of a class~I methanol masers are represented with red dots, other sources are represented with black dots.  The blue line shows the criteria developed by \citet{Chen+12} to identify BGPS sources likely to have an associated class~I methanol maser.}
\label{fig:classI_pred}
\end{figure}

When the different  classification models we developed using Data Set 3 are applied to the 8144 BGPS sources which have not been searched for class~I methanol maser emission a total of 739 sources are predicted to have an associated maser by one or more of the models, with 242 sources predicted by all four models (see Section~\ref{sec:classI} and Appendix~\ref{sec:predictions}).  Figure~12 of \citet{Chen+12} plots the integrated flux density against the beam averaged H$_2$ column density for Data Set 3 and shows that the maser associated sources are restricted to a limited range for these two predictor variables.  Figure~\ref{fig:classI_pred} shows the integrated flux density versus the beam averaged $H_2$ column density for the 8144 BGPS sources not observed by \citeauthor{Chen+12}.  Those sources for which one or more of the  classification models predict an associated class~I methanol maser are indicated with a red dot, with sources which no model predicts to have an associated class~I maser are indicated with a black dot.  Figure~\ref{fig:classI_pred} shows that there is a high level of agreement between the predictions of the  classification models and the empirical criteria developed by \citet{Chen+12}.  In total 1200 BGPS sources meet the criteria identified by \citet{Chen+12}, approximately a factor of two more than identified by any of the  classification models.  In their calculation of the beam averaged column density \citet{Chen+12} assumed a constant temperature of 20K for the dust clumps and used the 40 arcsecond flux density measurement as the intensity of the dust continuum emission.  This means that the calculated beam average column density is directly proportional to the BGPS 40 arcsecond flux density measurement.  The relationship derived by \citet{Chen+12} suggested this is the most important predictor variable for the presence (or otherwise) of class~I methanol maser emission towards these sources.  The results presented here also support this.

Ultimately, determining the relative accuracy of these classification models, and whether they are superior to directly derived criteria (such as those of \citet{Chen+12}) is to test them through future observations.  There are currently approximately 400 different class~I methanol maser sources which have been identified throughout the Galaxy \citep[see][and references therein]{Chen+13a}, so a search targeted towards the candidate BGPS sources we have identified is likely to significantly increase the number of known sources.

\section{Conclusions}

In this paper we present three major findings regarding the utilisation of different classification techniques on different size astronomical data sets.  1) For small data sets parametric methods (such as LDA and logistic regression perform better than random forests (a non-parametric method). 2) For larger data sets, random forests has the capability to out-perform the parametric methods trialled here. 3) In almost all cases, transforming the data to be closer to a normal distribution significantly increases the accuracy of  LDA.  In the case where using transformed data slightly decreased the accuracy of the model, the classification results were very similar.  Since the process of transforming data is relatively easy, this is a step that should be definitely employed if LDA is utilised.  This step has typically not been included when LDA has been applied to astronomical data used in past studies. 

Our results suggest that where there is very limited training information parametric models which can only predict based on simple combinations of the input variables are more accurate than non-parametric methods.  However, where there is more training data (such as Data Sets 1 and 3) non-parametric models can perform as well (likely better in some circumstances) than parametric techniques.   Our results for Data Set 3 show that random forests is comparable in accuracy to the parametric methods, rather than exceeding them as expected \citep[see][]{Breiman+01a}.

Frequently in astrophysics relationships are sought between two or three variables in the form of correlations between them, such as the radio:far-infrared correlation for galaxies, or colour-colour selection criteria for \ionhy regions.  In the past, this has often been because of limited numbers of predictor variables being available for large samples of data, however, this is now less of an issue.  Mathematical classification techniques such as those utilised here potentially offer significant improvements over simple correlation relationships, but the most appropriate technique to apply depends heavily on the nature of the data available and the goal of the investigation (e.g. detection prediction, physical understanding of relationship between variables).  Our models determined which predictor variables were important in the classification process, and for all three Data Sets our results agreed with the previous studies of \citet{Breen+07}, \citet{Ellingsen+10}, and \citet{Chen+12} respectively.  

For the specific goal of identifying millimetre dust clumps which are more likely to have an associated class I methanol maser, we find that on the basis of cross-validation tests and the predictions the models produce on the training data, both the non-parametric method of random forests and the parametric methods of logistic regression and LDA are well suited for the task of identifying likely targets for future searches.  242 sources out of the 8144 in Data Set 3, were predicted by all four of our techniques to have associated masers. The results of future searches for class I methanol masers towards BGPS sources will allow a direct test of each of the classification models and allow us to determine the validity of these conclusions.

\section*{Acknowledgements}

EMM undertook much of this work funded through a University of Tasmania Dean's Summer Research Scholarship.  Shari Breen is the recipient of an Australian Research Council DECRA Fellowship (project number DE130101270).  This research has made use of NASA's Astrophysics Data System Abstract Service.

\bibliographystyle{pasa-mnras}
\bibliography{pasaref}

\appendix

\section{Classification model predictions} \label{sec:predictions}

Table~\ref{tab:pred} summarises the predictions for each of the classification models for class~I methanol masers associated with Bolocam sources.

\begin{table*}
\caption{Bolocam Galactic Plane sources for which one or more of the mathematical classification models predicted the presence of an associated class~I methanol maser (probability of a maser $>$ 0.5).  The maser probability for each model is listed, those which exceed 0.5 are in bold type.  This list contains a total of 739 sources that were predicted to be masers by at least one of the four methods (242 of which were predicted by all methods), from a total of 8144 sources in version 1.0.1 of the Bolocam catalogue.}
\small
\begin{tabular}{clcccc} \hline
 \multicolumn{1}{c}{\bf Bolocam} & \multicolumn{1}{c}{\bf BGPS} & \multicolumn{1}{c}{\bf Random} & \multicolumn{1}{c}{\bf Logistic} & & \multicolumn{1}{c}{\bf Normalised} \\
 \multicolumn{1}{c}{\bf Catalogue \#} & \multicolumn{1}{c}{\bf name} & \multicolumn{1}{c}{\bf Forests} & \multicolumn{1}{c}{\bf Regression} & \multicolumn{1}{c}{\bf LDA} & \multicolumn{1}{c}{\bf LDA} \\ \hline \hline
   4 & G000.010$+$00.157 &    0.43   &    0.44   &    0.40   & \bf{0.60} \\ 
   5 & G000.016$-$00.017 & \bf{0.84} &    0.49   & \bf{0.64} &    0.42   \\ 
   7 & G000.020$+$00.033 & \bf{0.81} &    0.29   &    0.47   &    0.17   \\ 
   8 & G000.020$-$00.051 & \bf{0.92} & \bf{0.71} & \bf{0.75} &    0.45   \\ 
  18 & G000.052$+$00.027 & \bf{0.78} & \bf{0.72} & \bf{0.91} & \bf{0.64} \\ 
  20 & G000.054$-$00.209 & \bf{0.82} &    0.47   & \bf{0.77} &    0.31   \\
\hline
\end{tabular}
\label{tab:pred}
\end{table*}

\begin{table*}
\small
\begin{tabular}{clcccc} \hline
 \multicolumn{1}{c}{\bf Bolocam} & \multicolumn{1}{c}{\bf BGPS} & \multicolumn{1}{c}{\bf Random} & \multicolumn{1}{c}{\bf Logistic} & & \multicolumn{1}{c}{\bf Normalised} \\
 \multicolumn{1}{c}{\bf Catalogue \#} & \multicolumn{1}{c}{\bf name} & \multicolumn{1}{c}{\bf Forests} & \multicolumn{1}{c}{\bf Regression} & \multicolumn{1}{c}{\bf LDA} & \multicolumn{1}{c}{\bf LDA} \\ \hline \hline
 209 & G000.608$+$00.001 & \bf{0.83} &    0.40   &    0.08   &    0.13   \\ 
 211 & G000.610$-$00.057 & \bf{0.95} & \bf{0.99} & \bf{0.67} & \bf{0.96} \\ 
 218 & G000.630$-$00.095 & \bf{0.85} &    0.24   &    0.18   & \bf{0.73} \\ 
 222 & G000.648$+$00.027 & \bf{0.90} &    0.00   &    0.01   &    0.42   \\ 
 223 & G000.656$-$00.045 & \bf{0.77} & \bf{1.00} & \bf{1.00} & \bf{1.00} \\ 
 225 & G000.670$-$00.141 & \bf{0.88} &    0.01   &    0.21   &    0.30   \\
\hline
\end{tabular}
\end{table*}

\begin{table*}
\small
\begin{tabular}{clcccc} \hline
 \multicolumn{1}{c}{\bf Bolocam} & \multicolumn{1}{c}{\bf BGPS} & \multicolumn{1}{c}{\bf Random} & \multicolumn{1}{c}{\bf Logistic} & & \multicolumn{1}{c}{\bf Normalised} \\
 \multicolumn{1}{c}{\bf Catalogue \#} & \multicolumn{1}{c}{\bf name} & \multicolumn{1}{c}{\bf Forests} & \multicolumn{1}{c}{\bf Regression} & \multicolumn{1}{c}{\bf LDA} & \multicolumn{1}{c}{\bf LDA} \\ \hline \hline
 946 & G003.438$-$00.352 & \bf{0.95} & \bf{1.00} & \bf{0.98} & \bf{0.99} \\ 
 986 & G003.910$-$00.002 & \bf{0.61} &    0.33   &    0.10   &    0.24   \\ 
 987 & G003.932$-$00.008 & \bf{0.64} &    0.11   &    0.09   &    0.10   \\ 
1018 & G004.418$+$00.124 &    0.43   & \bf{0.51} &    0.13   & \bf{0.89} \\ 
1020 & G004.434$+$00.126 & \bf{0.91} & \bf{0.95} & \bf{0.59} & \bf{0.89} \\ 
1039 & G004.681$+$00.277 & \bf{0.67} & \bf{0.55} &    0.15   &    0.49   \\
\hline
\end{tabular}
\end{table*}

\begin{table*}
\small
\begin{tabular}{clcccc} \hline
 \multicolumn{1}{c}{\bf Bolocam} & \multicolumn{1}{c}{\bf BGPS} & \multicolumn{1}{c}{\bf Random} & \multicolumn{1}{c}{\bf Logistic} & & \multicolumn{1}{c}{\bf Normalised} \\
 \multicolumn{1}{c}{\bf Catalogue \#} & \multicolumn{1}{c}{\bf name} & \multicolumn{1}{c}{\bf Forests} & \multicolumn{1}{c}{\bf Regression} & \multicolumn{1}{c}{\bf LDA} & \multicolumn{1}{c}{\bf LDA} \\ \hline \hline
1659 & G011.947$-$00.036 & \bf{0.60} & \bf{0.90} &    0.34   & \bf{0.96} \\ 
1676 & G012.113$-$00.128 & \bf{0.65} &    0.22   &    0.08   &    0.17   \\ 
1683 & G012.209$-$00.104 & \bf{0.96} & \bf{1.00} & \bf{0.76} & \bf{0.99} \\ 
1684 & G012.215$-$00.118 & \bf{0.85} & \bf{0.89} &    0.01   & \bf{0.53} \\ 
1708 & G012.403$-$00.466 &    0.38   & \bf{0.52} &    0.41   &    0.48   \\ 
1710 & G012.419$+$00.506 & \bf{0.96} & \bf{1.00} & \bf{0.99} & \bf{0.98} \\
\hline
\end{tabular}
\end{table*}

\begin{table*}
\small
\begin{tabular}{clcccc} \hline
 \multicolumn{1}{c}{\bf Bolocam} & \multicolumn{1}{c}{\bf BGPS} & \multicolumn{1}{c}{\bf Random} & \multicolumn{1}{c}{\bf Logistic} & & \multicolumn{1}{c}{\bf Normalised} \\
 \multicolumn{1}{c}{\bf Catalogue \#} & \multicolumn{1}{c}{\bf name} & \multicolumn{1}{c}{\bf Forests} & \multicolumn{1}{c}{\bf Regression} & \multicolumn{1}{c}{\bf LDA} & \multicolumn{1}{c}{\bf LDA} \\ \hline \hline
2169 & G015.099$-$00.558 & \bf{0.55} &    0.43   &    0.30   &    0.33   \\ 
2170 & G015.099$-$00.600 & \bf{0.82} &    0.43   &    0.42   &    0.30   \\ 
2171 & G015.101$-$00.656 & \bf{0.90} & \bf{0.94} & \bf{0.64} & \bf{0.80} \\ 
2181 & G015.137$-$00.674 & \bf{0.85} & \bf{0.66} & \bf{0.73} &    0.35   \\ 
2184 & G015.153$-$00.660 & \bf{0.64} & \bf{0.67} &    0.17   &    0.30   \\ 
2189 & G015.182$-$00.158 &    0.26   &    0.47   &    0.33   & \bf{0.67} \\
\hline
\end{tabular}
\end{table*}

\begin{table*}
\small
\begin{tabular}{clcccc} \hline
 \multicolumn{1}{c}{\bf Bolocam} & \multicolumn{1}{c}{\bf BGPS} & \multicolumn{1}{c}{\bf Random} & \multicolumn{1}{c}{\bf Logistic} & & \multicolumn{1}{c}{\bf Normalised} \\
 \multicolumn{1}{c}{\bf Catalogue \#} & \multicolumn{1}{c}{\bf name} & \multicolumn{1}{c}{\bf Forests} & \multicolumn{1}{c}{\bf Regression} & \multicolumn{1}{c}{\bf LDA} & \multicolumn{1}{c}{\bf LDA} \\ \hline \hline
3029 & G023.274$-$00.212 & \bf{0.52} &    0.29   &    0.40   &    0.32   \\ 
3039 & G023.321$-$00.298 & \bf{0.50} &    0.44   & \bf{0.58} & \bf{0.55} \\ 
3053 & G023.368$-$00.290 & \bf{0.77} & \bf{0.83} &    0.26   & \bf{0.68} \\ 
3065 & G023.414$-$00.228 & \bf{0.89} &    0.03   & \bf{0.91} &    0.31   \\ 
3077 & G023.456$+$00.064 & \bf{0.69} & \bf{0.89} &    0.33   & \bf{0.85} \\ 
3078 & G023.456$-$00.018 &    0.24   &    0.05   &    0.06   & \bf{0.73} \\
\hline
\end{tabular}
\end{table*}

\begin{table*}
\small
\begin{tabular}{clcccc} \hline
 \multicolumn{1}{c}{\bf Bolocam} & \multicolumn{1}{c}{\bf BGPS} & \multicolumn{1}{c}{\bf Random} & \multicolumn{1}{c}{\bf Logistic} & & \multicolumn{1}{c}{\bf Normalised} \\
 \multicolumn{1}{c}{\bf Catalogue \#} & \multicolumn{1}{c}{\bf name} & \multicolumn{1}{c}{\bf Forests} & \multicolumn{1}{c}{\bf Regression} & \multicolumn{1}{c}{\bf LDA} & \multicolumn{1}{c}{\bf LDA} \\ \hline \hline
4048 & G028.811$+$00.169 & \bf{0.63} &    0.26   & \bf{0.61} &    0.40   \\ 
4049 & G028.817$+$00.363 &    0.34   & \bf{0.81} &    0.18   & \bf{0.87} \\ 
4055 & G028.831$-$00.255 & \bf{0.86} & \bf{1.00} & \bf{0.66} & \bf{0.99} \\ 
4061 & G028.863$+$00.065 & \bf{0.73} & \bf{0.98} & \bf{0.60} & \bf{0.94} \\ 
4063 & G028.881$-$00.025 & \bf{0.64} &    0.28   &    0.12   &    0.19   \\ 
4121 & G029.225$+$00.023 & \bf{0.54} &    0.36   &    0.10   &    0.15   \\
\hline
\end{tabular}
\end{table*}

\begin{table*}
\small
\begin{tabular}{clcccc} \hline
 \multicolumn{1}{c}{\bf Bolocam} & \multicolumn{1}{c}{\bf BGPS} & \multicolumn{1}{c}{\bf Random} & \multicolumn{1}{c}{\bf Logistic} & & \multicolumn{1}{c}{\bf Normalised} \\
 \multicolumn{1}{c}{\bf Catalogue \#} & \multicolumn{1}{c}{\bf name} & \multicolumn{1}{c}{\bf Forests} & \multicolumn{1}{c}{\bf Regression} & \multicolumn{1}{c}{\bf LDA} & \multicolumn{1}{c}{\bf LDA} \\ \hline \hline
5053 & G032.798$+$00.193 & \bf{0.93} & \bf{1.00} & \bf{0.99} & \bf{1.00} \\ 
5057 & G032.820$-$00.329 &    0.34   & \bf{0.62} &    0.30   & \bf{0.80} \\ 
5120 & G033.133$-$00.091 & \bf{0.60} & \bf{0.99} & \bf{0.85} & \bf{0.99} \\ 
5171 & G033.414$-$00.002 & \bf{0.58} &    0.24   &    0.32   &    0.19   \\ 
5229 & G033.652$-$00.025 & \bf{0.55} &    0.18   &    0.46   &    0.22   \\ 
5263 & G033.810$-$00.187 &    0.20   &    0.49   &    0.33   & \bf{0.51} \\
\hline
\end{tabular}
\end{table*}

\begin{table*}
\small
\begin{tabular}{clcccc} \hline
 \multicolumn{1}{c}{\bf Bolocam} & \multicolumn{1}{c}{\bf BGPS} & \multicolumn{1}{c}{\bf Random} & \multicolumn{1}{c}{\bf Logistic} & & \multicolumn{1}{c}{\bf Normalised} \\
 \multicolumn{1}{c}{\bf Catalogue \#} & \multicolumn{1}{c}{\bf name} & \multicolumn{1}{c}{\bf Forests} & \multicolumn{1}{c}{\bf Regression} & \multicolumn{1}{c}{\bf LDA} & \multicolumn{1}{c}{\bf LDA} \\ \hline \hline
6312 & G049.192$-$00.336 & \bf{0.93} & \bf{0.88} & \bf{0.74} & \bf{0.94} \\ 
6313 & G049.210$-$00.342 & \bf{0.89} & \bf{0.97} & \bf{1.00} & \bf{0.92} \\ 
6321 & G049.264$+$00.312 & \bf{0.63} &    0.09   &    0.10   &    0.07   \\ 
6334 & G049.367$-$00.302 & \bf{0.87} & \bf{1.00} & \bf{1.00} & \bf{0.98} \\ 
6336 & G049.371$-$00.350 & \bf{0.70} &    0.25   & \bf{0.52} &    0.20   \\ 
6337 & G049.375$-$00.262 & \bf{0.87} & \bf{0.84} & \bf{0.66} & \bf{0.55} \\
\hline
\end{tabular}
\end{table*}

\begin{table*}
\small
\begin{tabular}{clcccc} \hline
 \multicolumn{1}{c}{\bf Bolocam} & \multicolumn{1}{c}{\bf BGPS} & \multicolumn{1}{c}{\bf Random} & \multicolumn{1}{c}{\bf Logistic} & & \multicolumn{1}{c}{\bf Normalised} \\
 \multicolumn{1}{c}{\bf Catalogue \#} & \multicolumn{1}{c}{\bf name} & \multicolumn{1}{c}{\bf Forests} & \multicolumn{1}{c}{\bf Regression} & \multicolumn{1}{c}{\bf LDA} & \multicolumn{1}{c}{\bf LDA} \\ \hline \hline
6747 & G080.829$+$00.568 & \bf{0.74} & \bf{0.69} & \bf{0.55} & \bf{0.61} \\ 
6753 & G080.863$+$00.384 & \bf{0.79} & \bf{0.96} & \bf{0.69} & \bf{0.94} \\ 
6754 & G080.864$+$00.422 & \bf{0.79} & \bf{0.98} & \bf{0.90} & \bf{0.97} \\ 
6762 & G080.941$-$00.126 & \bf{0.69} & \bf{0.61} & \bf{0.75} & \bf{0.72} \\ 
6765 & G080.954$-$00.154 & \bf{0.61} &    0.45   &    0.19   &    0.34   \\ 
6788 & G081.117$-$00.140 & \bf{0.72} &    0.28   & \bf{0.61} &    0.18   \\
\hline
\end{tabular}
\end{table*}

\begin{table*}
\small
\begin{tabular}{clcccc} \hline
 \multicolumn{1}{c}{\bf Bolocam} & \multicolumn{1}{c}{\bf BGPS} & \multicolumn{1}{c}{\bf Random} & \multicolumn{1}{c}{\bf Logistic} & & \multicolumn{1}{c}{\bf Normalised} \\
 \multicolumn{1}{c}{\bf Catalogue \#} & \multicolumn{1}{c}{\bf name} & \multicolumn{1}{c}{\bf Forests} & \multicolumn{1}{c}{\bf Regression} & \multicolumn{1}{c}{\bf LDA} & \multicolumn{1}{c}{\bf LDA} \\ \hline \hline
7394 & G134.211$+$00.729 & \bf{0.67} &    0.04   &    0.16   &    0.26   \\ 
7396 & G134.218$+$00.787 & \bf{0.73} &    0.26   &    0.30   &    0.33   \\ 
7456 & G138.295$+$01.556 & \bf{0.91} & \bf{0.96} & \bf{0.82} & \bf{0.85} \\ 
7459 & G138.503$+$01.646 & \bf{0.53} & \bf{0.62} &    0.38   & \bf{0.60} \\ 
7460 & G188.792$+$01.027 & \bf{0.75} & \bf{0.56} & \bf{0.79} & \bf{0.52} \\ 
7461 & G188.948$+$00.883 & \bf{0.81} & \bf{1.00} & \bf{1.00} & \bf{0.99} \\
\hline
\end{tabular}
\end{table*}

\begin{table*}
\small
\begin{tabular}{clcccc} \hline
 \multicolumn{1}{c}{\bf Bolocam} & \multicolumn{1}{c}{\bf BGPS} & \multicolumn{1}{c}{\bf Random} & \multicolumn{1}{c}{\bf Logistic} & & \multicolumn{1}{c}{\bf Normalised} \\
 \multicolumn{1}{c}{\bf Catalogue \#} & \multicolumn{1}{c}{\bf name} & \multicolumn{1}{c}{\bf Forests} & \multicolumn{1}{c}{\bf Regression} & \multicolumn{1}{c}{\bf LDA} & \multicolumn{1}{c}{\bf LDA} \\ \hline \hline
7763 & G353.362$-$00.088 & \bf{0.50} &    0.22   &    0.19   &    0.38   \\ 
7764 & G353.365$-$00.166 & \bf{0.81} & \bf{0.99} &    0.46   & \bf{0.87} \\ 
7765 & G353.367$-$00.336 & \bf{0.72} &    0.26   &    0.34   & \bf{0.56} \\ 
7767 & G353.384$-$00.336 & \bf{0.66} & \bf{0.63} &    0.47   & \bf{0.75} \\ 
7770 & G353.400$-$00.070 & \bf{0.89} & \bf{1.00} &    0.45   & \bf{0.89} \\ 
7771 & G353.412$-$00.360 & \bf{0.91} & \bf{1.00} & \bf{1.00} & \bf{1.00} \\
\hline
\end{tabular}
\end{table*}

\begin{table*}
\small
\begin{tabular}{clcccc} \hline
 \multicolumn{1}{c}{\bf Bolocam} & \multicolumn{1}{c}{\bf BGPS} & \multicolumn{1}{c}{\bf Random} & \multicolumn{1}{c}{\bf Logistic} & & \multicolumn{1}{c}{\bf Normalised} \\
 \multicolumn{1}{c}{\bf Catalogue \#} & \multicolumn{1}{c}{\bf name} & \multicolumn{1}{c}{\bf Forests} & \multicolumn{1}{c}{\bf Regression} & \multicolumn{1}{c}{\bf LDA} & \multicolumn{1}{c}{\bf LDA} \\ \hline \hline
8288 & G359.713$+$00.045 & \bf{0.53} &    0.20   &    0.42   &    0.11   \\ 
8289 & G359.716$-$00.375 & \bf{0.82} & \bf{0.74} & \bf{0.54} & \bf{0.74} \\ 
8299 & G359.752$+$00.037 & \bf{0.51} &    0.07   &    0.36   &    0.03   \\ 
8319 & G359.822$+$00.029 &    0.49   &    0.05   & \bf{0.51} &    0.12   \\ 
8329 & G359.864$+$00.019 & \bf{0.53} &    0.14   &    0.35   &    0.22   \\ 
8332 & G359.867$-$00.085 & \bf{0.89} &    0.16   &    0.25   & \bf{0.95} \\ 
8335 & G359.891$-$00.071 & \bf{0.80} & \bf{0.98} & \bf{1.00} & \bf{0.94} \\ 
8337 & G359.900$+$00.015 & \bf{0.70} &    0.15   &    0.23   &    0.11   \\ 
8338 & G359.906$+$00.041 & \bf{0.73} &    0.43   &    0.15   &    0.45   \\ 
8342 & G359.912$-$00.305 &    0.10   &    0.32   &    0.49   & \bf{0.56} \\ 
8345 & G359.920$+$00.025 & \bf{0.75} &    0.32   &    0.13   &    0.16   \\ 
8353 & G359.944$+$00.171 & \bf{0.87} & \bf{0.82} &    0.49   & \bf{0.77} \\ 
8354 & G359.946$+$00.153 & \bf{0.94} & \bf{1.00} &    0.39   & \bf{0.95} \\ 
8355 & G359.946$-$00.047 & \bf{0.91} & \bf{1.00} & \bf{1.00} & \bf{1.00} \\ 
8356 & G359.947$+$00.023 & \bf{0.78} &    0.14   &    0.45   &    0.08   \\ 
8361 & G359.971$-$00.459 & \bf{0.59} & \bf{0.99} & \bf{0.83} & \bf{0.97} \\ 
8366 & G359.978$-$00.071 & \bf{0.85} &    0.00   & \bf{1.00} & \bf{0.84} \\ 
8367 & G359.985$+$00.023 & \bf{0.65} &    0.11   &    0.43   &    0.10   \\ 
8370 & G359.994$+$00.107 & \bf{0.60} &    0.11   &    0.26   &    0.04   \\ \hline
\end{tabular}
\end{table*}

\end{document}